\documentclass[aps,prd,superscriptaddress,nofootinbib,11pt]{revtex4}
\usepackage[english]{babel}
\usepackage[utf8]{inputenc}
\usepackage{graphicx}   
\usepackage{slashed}
\usepackage{epstopdf}
\usepackage{verbatim}   
\usepackage{color}      
\usepackage{subfigure}  
\usepackage{multirow}
\usepackage{hyperref}   
\usepackage{float}
\usepackage{epsfig,rotating}
\usepackage{amsmath,amssymb}
\usepackage{dsfont}
\restylefloat{table}
\numberwithin{equation}{section}

\newcommand{\vx}{\vec{x}}
\newcommand{\vp}{\vec{p}}
\newcommand{\vq}{\vec{q}}
\newcommand{\vk}{\vec{k}}

\newcommand{\be}{\begin{equation}}
\newcommand{\ee}{\end{equation}}
\newcommand{\bea}{\begin{eqnarray}}
\newcommand{\eea}{\end{eqnarray}}

\newcommand{\ket}[1]{|#1\rangle}
\newcommand{\bra}[1]{\langle#1|}
\newcommand{\order}[1]{\mathcal{O}(#1)}
\newcommand{\Dk}{|\Delta_k|}

\newcommand{\mode}{e^{-i\int^\eta W_k(\eta')\,d\eta'}}

\begin{document}
\title{Particle decay in post inflationary cosmology.}

\author{Nathan Herring}
\email{nmh48@pitt.edu} \affiliation{Department of Physics and
Astronomy, University of Pittsburgh, Pittsburgh, PA 15260}
\author{Brian Pardo}
\email{bap100@pitt.edu} \affiliation{Department of Physics and
Astronomy, University of Pittsburgh, Pittsburgh, PA 15260}
\author{Daniel Boyanovsky}
\email{boyan@pitt.edu} \affiliation{Department of Physics and
Astronomy, University of Pittsburgh, Pittsburgh, PA 15260}
\author{Andrew R. Zentner}
\email{zentner@pitt.edu}
\affiliation{Department of Physics and
Astronomy, University of Pittsburgh, Pittsburgh, PA 15260}
\affiliation{Pittsburgh Particle Physics, Astrophysics, and Cosmology Center (Pitt PACC)}
\date{\today}

\begin{abstract}
We study a scalar particle  of mas $m_1$ decaying into two particles of mass $m_2$ during the radiation and matter dominated epochs of a
standard cosmological model. An adiabatic approximation is introduced that is
valid for degrees of freedom with typical wavelengths much smaller
than the particle horizon ($\propto$~Hubble radius) at a given time.
We implement a non-perturbative method that includes the cosmological
expansion and obtain a \emph{cosmological Fermi's Golden Rule} that
enables one to compute the \emph{decay law} of the parent particle  with mass
$m_1$, along with the build up of the population of daughter particles with
mass $m_2$. The survival probability of the decaying particle is
$P(t)=e^{-\widetilde{\Gamma}_k(t)\,t}$
with $\widetilde{\Gamma}_k(t)$ being an \emph{effective momentum and time dependent decay rate}.
It  features a transition time scale $t_{nr}$ between the relativistic
and non-relativistic regimes and for $k \neq 0$ is \emph{always} smaller than the analogous
rate in Minkowski spacetime,    as a consequence of
(local) time dilation and the cosmological redshift.
For $t \ll t_{nr}$  the decay law is a ``stretched exponential''
$P(t) = e^{-(t/t^*)^{3/2}}$, whereas for the non-relativistic stage with
$t \gg t_{nr}$, we find $P(t) = e^{-\Gamma_0 t}\,(t/t_{nr})^{\Gamma_0\,t_{nr}/2}$, with $\Gamma_0$ the Minkowski space time decay width at rest.  The Hubble time scale $\propto 1/H(t)$ introduces an energy
uncertainty $\Delta E \sim H(t)$
which relaxes the constraints of kinematic thresholds.
This opens new decay channels into \emph{heavier particles} for
$2\pi E_k(t) H(t) \gg 4m^2_2-m^2_1$, with $E_k(t)$ the (local)
comoving energy of the decaying particle.
As the expansion proceeds this channel closes and
the usual two particle threshold  restricts the decay kinematics.

\end{abstract}


\maketitle

\section{Introduction}

Particle decay is an ubiquitous process that has profound implications in cosmology, for baryogenesis \cite{decaybaryo,decaybaryo2}, leptogenesis \cite{wimpdmlepto,leptodecay}, CP violating decays \cite{roulet}, big bang nucleosynthesis (BBN) \cite{zw2002,bbnconstdecay,bbnlonglivedecay,fieldsbbn,ishidabbn,serpicobbn,pospelovbbn,poulinbbn,salvatibbn}, and dark matter (DM) where large scale structure and supernova Ia luminosity distances constrain the lifetimes of potential, long-lived candidates \cite{zw2002,zentner,zentnerlyalfa,koush,dmdecayconst,dmconst2}. Most analyses of particle decay in cosmology use decay rates obtained from S-matrix theory in Minkowski spacetime. In this formulation, the decay rate is obtained from the total transition probability from a state prepared in the infinite past (in) to final states in the infinite future (out). Dividing this probability by the total time elapsed enables one to extract a transition probability per unit time. Energy conservation emerging in the infinite time limit yields kinematic constraints (thresholds) for decay processes.

The decay rate so defined, and calculated, is an input in analyses of cosmological processes. In an expanding cosmology with a time-dependent gravitational background, there is no global time-like Killing vector; therefore, particle energy is not manifestly conserved, even in spatially flat Friedmann-Robertson-Walker (FRW) cosmologies, which do supply spatial momentum conservation. Early studies of quantum field theory in curved space-time revealed a wealth of unexpected novel phenomena, such as particle production from cosmological expansion \cite{parker,zelstaro,fulling,birford,bunch,birrell,fullbook,mukhabook,parkerbook,parfull} and the possibility of processes that are forbidden in Minkowski space time as a consequence of energy/momentum conservation. Pioneering investigations of interacting quantum fields in expanding cosmologies generalized the S-matrix formulation for in-out states in Minkowski spacetimes for model expansion histories. Self-interacting quantized fields were studied with a focus on renormalization aspects and contributions from pair production to the energy momentum tensor \cite{birford,bunch}. The decay of a massive particle into two massless particles conformally coupled to gravity was studied in Ref.~\cite{spangdecay} within the context of in-out S-matrix for simple cosmological space times. Particle decay in inflationary cosmology (near de Sitter space-time) was studied in Refs.~\cite{boydecay,mosca}, revealing surprising phenomena, such as a quantum of a massive field decaying into two (or more) quanta of the \emph{same} field. The lack of a global, time-like Killing vector, and the concomitant absence of energy conservation, enables such remarkable processes that are forbidden in Minkowski spacetime. More recently, the methods introduced in Ref.~\cite{spangdecay} were adapted to study the decay of a massive particle into two conformally massless particles in radiation and ``stiff'' matter dominated cosmology, focusing on extracting a decay rate for \emph{zero momentum} \cite{vilja}. The results of Ref.~\cite{vilja} approach those of Minkowski spacetime asymptotically in the long-time limit.

 \vspace{1mm}

 \textbf{Motivation, goals and summary of results}.
The importance and wide range of phenomenological consequences of particle decay in cosmology motivate us to study this process within the realm of the standard post inflationary cosmology,  during the radiation and matter dominated eras. Our goal is to obtain and implement a quantum field theory framework that includes consistently the cosmological expansion and that can be applied to the various interactions and fields of the standard model and beyond.

 \vspace{1mm}
 \textbf{Brief summary of  results:} We combine a physically motivated adiabatic expansion with a non-perturbative method that is the \emph{quantum field theoretical} version of the Wigner-Weisskopf theory of atomic line-widths\cite{ww} ubiquitous in quantum optics \cite{book1}. This method is manifestly unitary, and has been implemented in both Minkowski spacetime and inflationary cosmology \cite{boyww,boycosww}, and provides a systematic framework to obtain the \emph{decay law} of the parent along with the production probability of the daughter particles.  One of our main results, to leading order in this adiabatic expansion, is a \emph{cosmological Fermi's Golden Rule} wherein the particle horizon (proportional to the Hubble time) determines an uncertainty in the (local) comoving energy. We find that the parent survival probability may be written in terms of an \emph{effective time-dependent decay rate} which includes the effects of (local) time dilation and cosmological redshift, resulting in a delayed decay. This effective rate depends crucially on a transition time, $t_{nr}$, between the relativistic and non-relativistic regimes of the parent particle, and is always \emph{smaller} than that in Minkowski spacetime, becoming equal only in the limit of a parent particle always at rest in the comoving frame. An unexpected consequence of the cosmological expansion is that the uncertainty implied by the particle horizon opens new decay channels to particles \emph{heavier} than the parent. As the expansion proceeds this channel closes and the usual kinematic thresholds constrain the phase space for the decay process. While in this study we focus on the radiation dominated (RD) era, our results can be simply extended to the subsequent matter dominated (MD) and dark energy dominated eras. In appendix (\ref{app:Mink}) we implement the Wigner-Weisskopf method in Minkowski spacetime to provide a basis of comparison which will enable us to highlight the major differences with the cosmological setting.

\section{The standard post-inflationary cosmology}

We focus on the decay of particles in the post-inflationary universe, described by a   spatially flat (FRW) cosmology  with the metric in comoving coordinates given by
\be  g_{\mu \nu} = \textrm{diag}(1, -a^2, -a^2, -a^2) \,. \label{frwmetric}\ee The standard cosmology post-inflation is described by  three distinct stages: radiation (RD), matter (MD) and dark energy (DE) domination; we model the latter by a cosmological constant. Friedmann's equation is
\be \Big(\frac{\dot{a}}{a}\Big)^2 = H^2(t) = H^2_0\,\Bigg[ \frac{\Omega_M}{a^3(t)}+ \frac{\Omega_R}{a^4(t)}+ \Omega_\Lambda \Bigg] \,,\label{Hubble} \ee
where the scale factor is normalized to $a_0=a(t_0)=1$ today. We take as representative the following  values of the parameters \cite{wmap,spergel,planck}:
\be H_0 = 1.5\times 10^{-42}\,\mathrm{GeV}~~;~~ \Omega_M = 0.308~~;~~ \Omega_R = 5\times 10^{-5}~~;~~ \Omega_\Lambda = 0.692 \,.\label{cosmopars}\ee
It is convenient to pass from ``comoving time,'' $t$,   to conformal time $\eta$ with $d\eta = dt/a(t)$, in terms of which the metric becomes ($a\equiv a(\eta)$)
\be  g_{\mu \nu} = \textrm{diag}(a^2,-a^2,-a^2,-a^2) \, . \label{conformalmetric} \ee
With (${\,}^{\,'} \equiv \frac{d}{d\eta}$) we find
\be a'(\eta) = H_0\,\sqrt{\Omega_M}\,\Big[r+a+s\,a^4\Big]^{1/2}\,,\label{dera}\ee with
\be r= \frac{\Omega_R}{\Omega_M} \simeq 1.66\,\times 10^{-4}~~;~~ s = \frac{\Omega_\Lambda}{\Omega_M} \simeq 2.25 \,.\label{rands}\ee
Hence the different stages of cosmological evolution, namely radiation domination (RD), matter domination (MD), and dark energy domination (DE), are characterized by
\be a\ll r \Rightarrow \text{RD}~~;~~ r \ll a \lesssim 0.76 \Rightarrow \text{MD} ~~;~~ a > 0.76 \Rightarrow \text{DE} \,. \label{stages}\ee
In the standard cosmological picture and the majority of the most well-studied variants, most of the interesting particle physics processes occur during the RD era and so we focus most of our attention on this epoch; however, we also contemplate the possibility of long-lived dark matter particles that would decay on time scales of the order of $1/H_0$. The RD and MD epochs cover approximately half of the age of the Universe and during these stages the evolution of the scale factor can be written as
\be a(\eta)= H_R   ~\eta + \frac{H^2_M}{4}  ~\eta^2 ~~;~~ H_R= H_0\,\sqrt{\Omega_R},~~;~~H_M= H_0\,\sqrt{\Omega_M}\,,  \label{aofetarm}\ee
which facilitates the  explicit analytical study of the decay laws.
In turn, the conformal time at a given scale factor $a$ is given by
\be \eta(a) = \frac{2\,\sqrt{r}}{H_M}\,\Bigg[\sqrt{1 + \frac{a}{r}}-1 \Bigg]\,. \label{etaofa}\ee During the (RD) stage   the relation between conformal and comoving time is given by \be \eta = \Big( \frac{2\,t}{H_R}\Big)^{\frac{1}{2}} \Rightarrow a(t) = \Big[ 2\,t H_R\Big]^{\frac{1}{2}}\,, \label{etaoft} \ee a result that will prove useful in the study of the decay law during this stage.

\section{The model:}

We consider  two interacting scalar fields $\phi_1,\phi_2$ in  the FRW cosmology determined by the metric (\ref{frwmetric}),  with action given by
\be  A   =    \int d^4 x \sqrt{|g|} \Bigg\{\frac{1}{2} g^{\mu\nu}\,\partial_\mu \phi_1 \partial_\nu \phi_1-\frac{1}{2} \big[m^2_1 +\xi_1\,R\big]\phi^2_1 + \frac{1}{2} g^{\mu\nu}\,\partial_\mu \phi_2 \partial_\nu \phi_2-\frac{1}{2} \big[m^2_2 +\xi_2\,R\big]\phi^2_2   -   \lambda \phi_1 :\phi^2_2:  \Bigg\}
\label{action}\ee where
\be R= 6\Big[\frac{\ddot{a}}{a}+\Big(\frac{\dot{a}}{a} \Big)^2 \Big] \label{ricci}\ee is the Ricci scalar,  and $\xi_{1,2}$ are couplings to gravity, with $\xi=0, 1/6$ corresponding to minimal or conformal coupling, respectively. We identify $\phi_1$ as the field associated with the decaying (``parent") particle, and $\phi_2$ as that of the decay product (``daughter") particles.

Expressing the action of Eq.~(\ref{action}) in terms of the comoving spatial coordinates and the conformal time, while rescaling the fields as
\be \phi_{1,2}(\vec{x},t) = \frac{\chi_{1,2}(\vec{x},\eta)}{a(\eta)} ~~;~~ a(\eta) = a(t(\eta))\,, \label{rescale}\ee
yields
\be A= \int d^3x \, d\eta  \biggl\{\sum_{j=1,2}\Bigl[\tfrac{1}{2}  \,\Big(\frac{d\chi_j}{d\eta} \Big)^2 -\tfrac{1}{2} \,\bigl( \nabla \chi_j \bigr)^2 -\tfrac{1}{2} \chi^2_j\, \mathcal{M}^2_j(\eta) \Bigr] -   \lambda \,a(\eta)\,\chi_1 \,:\chi^2_2:\,  \biggr\}\,
\label{conformalaction}\ee
neglecting surface terms as usual, where
\be \mathcal{M}^2_{j}(\eta) = m^2_j\,a^2(\eta)- \frac{a''}{a}\,(1-6\xi_j) ~~;~~j=1,2 \,. \label{massconformal}\ee
For the standard cosmology, using (\ref{dera})
\be \frac{a''}{a} = \frac{H^2_M}{2\,a(\eta)} \Big[1+4s a^3(\eta) \Big]\,.  \label{Ricci} \ee

\vspace{2mm}

\textbf{Quantization:} We begin with the quantization of free fields \cite{birrell,fullbook,mukhabook,parkerbook,birford} ($\lambda =0$) as a prelude to the interacting theory. The Heisenberg equations of motion for the conformally rescaled fields in conformal time are
\be \frac{d^2}{d\eta^2}\,\chi_{j}(\vec{x},\eta) - \nabla^2 \chi_j(\vec{x},\eta) + \mathcal{M}^2_j(\eta)\,\chi_j(\vec{x},\eta) = 0 ~~;~~j=1,2 \,.\label{Eom}\ee It is convenient to consider the spatial Fourier transform in a comoving volume $V$, namely,
\be \chi(\vec{x},\eta) = \frac{1}{\sqrt{V}}\,\smash[b]{\sum_{\vec{k}}} \chi_{\vk}(\eta)\,e^{-i\vk\cdot\vx}\,,\label{FT}\ee leading to
\be \frac{d^2}{d\eta^2}\,\chi_{\vk} (\eta) +\Big[\omega^2_k(\eta)-\frac{a''}{a}\,(1-6\xi_j) \Big]\, \chi_{\vk} (\vec{k},\eta)   = 0 ~~;~~\omega^2_k(\eta) = k^2+m^2_j\,a^2(\eta) \,,\label{EomFT}\ee for either field, respectively.

Although solutions of (\ref{EomFT}) can be found for separate stages or model expansion histories\cite{vilja}, solving for the exact mode functions     for the standard cosmology with the different stages,  even when neglecting the term with $a''/a$,  is not  feasible. Instead we   focus on obtaining approximate solutions in an adiabatic expansion\cite{birrell,fullbook,mukhabook,parkerbook,birford,dunne,wini} that relies on a separation of time scales between those of the particle physics process and that of the cosmological expansion. As an example, let us consider a  physically motivated setting wherein the decaying particle   has been produced (``born'')  early during the radiation dominated stage
by the decay of heavier particle states at the Grand Unification (GUT) scale  $\simeq 10^{15}\,\mathrm{GeV}$. Assuming that the
mass of the (DM) particle is much smaller than this scale, the production process will endow the (DM)  particle with a \emph{physical} momentum $k_p(\eta) = k/a(\eta) \simeq 10^{15}\,\mathrm{GeV}$ with $k$ being the \emph{comoving} momentum. If the environmental temperature of the plasma is $T \simeq T_{\text{GUT}} \simeq 10^{15}\,\mathrm{GeV}$ and neglecting the processes that reheat the photon bath by entropy injection from massive degrees of freedom, then
$T_{\text{GUT}} \simeq T_\text{CMB}/a(\eta_i)$ implying that the scale factor at the GUT scale $a(\eta_i) \simeq 10^{-28}$. In turn this estimate implies that the \emph{comoving} wavevector $k$ with which the (DM) is produced is $k \simeq 10^{-13}\,\mathrm{GeV}$.

The result (\ref{Ricci}) suggests that when considering initial conditions at the GUT scale (or below)  corresponding to  $a(\eta_i) \geq 10^{-28}$ the term $a''/a$ in (\ref{EomFT}) can be neglected for $\omega_k(\eta_i) \gg 10^{-30}\,\mathrm{GeV}$ for scalar fields minimally coupled to gravity (or for any $|\xi_j| \simeq \mathcal{O}(1)$), since $\omega^2_k(\eta_i) \gg \frac{H^2_m}{2a(\eta_i)}$.  This condition is most certainly realized for particles produced from processes at the GUT scale, since as argued above such processes would yield comoving wavectors $k \simeq 10^{-13} \,\mathrm{GeV}$, hence $\omega_k(\eta_i) \geq 10^{-13}\,\mathrm{GeV}$ for (DM) particles (or daughters) with masses below the GUT scale. Therefore under these conditions we can safely ignore the term with $a''/a$ in (\ref{EomFT}). Below (see eqn. (\ref{addot}) and following comments) we show explicitly that this term is of second order in the adiabatic expansion and can be ignored to leading order.  The mode equations  (\ref{EomFT}) now become
\be \frac{d^2}{d\eta^2}\,\chi_{\vk} (\eta) + \omega^2_k(\eta)\, \chi_{\vk} (\eta)   = 0 \,. \label{eqnofmot}\ee

Field quantization is achieved by writing
\be \chi_{\vk} = a_{\vk}\,g_k(\eta) + a^\dagger_{-\vk}\, g^*_k(\eta) \,, \label{quant}\ee where the mode functions $g_k(\eta)$ obey the equation of motion
\be \frac{d^2}{d\eta^2}\,g_k (\eta) + \omega^2_k(\eta)\, g_k (\eta)   = 0 \,, \label{geqnofmot}\ee with the Wronskian condition
\be g^{\,'}_k(\eta)g^*_k(\eta) - g^{*\,'}_k(\eta) g_k(\eta) = -i \,  \label{wron}\ee so that the annihilation $a_{\vk}$ and creation $a^\dagger_{\vk}$ operators are \emph{time independent} and obey the canonical commutation relations $[a_{\vk},a^{\dagger}_{\vk'}] = \delta_{\vk,\vk'}$.

Writing the solution of this equation in the WKB form\cite{birrell,fullbook,mukhabook,parkerbook,birford}
\be g_k (\eta) = \frac{e^{-i\,\int^{\eta}_{\eta_i}\,W_k(\eta')\,d\eta'}}{\sqrt{2\,W_k(\eta)}} \,, \label{WKB}\ee and inserting this ansatz into (\ref{eqnofmot}) it follows that $W_k(\eta)$ must be a solution of the equation\cite{birrell}
\be W^2_k(\eta)= \omega^2_k(\eta)- \frac{1}{2}\bigg[\frac{W^{''}_k(\eta)}{W_k(\eta)} - \frac{3}{2}\,\bigg(\frac{W^{'}_k(\eta)}{W_k(\eta)}\bigg)^2 \bigg]\,.  \label{WKBsol} \ee

This equation can be solved in an \emph{adiabatic expansion}
\be W^2_k(\eta)= \omega^2_k(\eta) \,\bigg[1 - \frac{1}{2}\,\frac{\omega^{''}_k(\eta)}{\omega^3_k(\eta)}+
\frac{3}{4}\,\bigg( \frac{\omega^{'}_k(\eta)}{\omega^2_k(\eta)}\bigg)^2 +\cdots  \bigg] \,.\label{adexp}\ee We refer to terms that feature $n$-derivatives of $\omega_k(\eta)$ as of n-th adiabatic order. The nature and reliability of the adiabatic expansion is revealed by considering the term of first adiabatic order for generic mass $m$:
\be \frac{\omega^{'}_k(\eta)}{\omega^2_k(\eta)} = \frac{m^2\, a(\eta) a^{'}(\eta)}{\Big[ k^2 + m^2\,a^2(\eta) \Big]^{3/2}}\,, \label{firstordad}\ee this is most easily recognized in \emph{comoving} time $t$, introducing the \emph{local} energy $E_k(t)$ and Lorentz factor $\gamma_k(t)$ measured by a comoving observer in terms of the \emph{physical} momentum $k_p(t) = k/a(t)$
\bea E_k(t) & = &  \sqrt{k^2_p(t)+m^2}   \label{comoener}\\
 \gamma_k(t) & = & \frac{E_k(t)}{m} \,,\label{lorfac} \eea  and the Hubble expansion rate
 $H(t) = \frac{\dot{a}(t)}{a(t)} = a^{'}/a^2 $.    In terms of these variables, the first order adiabatic ratio  (\ref{firstordad}) becomes
 \be \frac{\omega^{'}_k(\eta)}{\omega^2_k(\eta)} = \frac{H(t)}{\gamma^2_k(t)\,E_k(t)}\,.  \label{adratio}\ee

 In similar fashion the higher order terms in the adiabatic expansion can be constructed as well:
\begin{align}
\frac{\omega^{''}_k(\eta)}{\omega_k^3(\eta)} &= \frac{m^2 \big((a^{'}(\eta))^2+a(\eta)a^{''}(\eta)\big)}{\omega_k^4(\eta)} - \frac{m^4a^2(\eta)(a^{'}(\eta))^2}{\omega_k^6(\eta)} \nonumber\\
&= \frac{1}{\gamma^2_k(t)}\Big(\frac{R(t)}{6 E^2_k(t)}+\frac{H^2(t)}{E_k^2(t)}\Big)-\frac{H^2(t)}{\gamma^4_k(t)E_k^2(t)}\,, \label{secordad}
\end{align}
where $R(t)$ is the Ricci scalar (\ref{ricci}).  Consequently, (\ref{adexp}) takes the form:
\begin{equation}
W^2_k(t) = a^2(t)E^2_k(t)\Big[1-\frac{1}{2\gamma^2_k(t)}\Big(\frac{R(t)}{6E^2_k(t)}+\frac{H^2(t)}{E_k^2(t)}\Big) + \frac{5}{4}\frac{H^2(t)}{\gamma^4_k(t)E_k^2(t)} +\cdots \Big]\,.
\end{equation}

Consider that the decaying (parent) particle is produced during the  radiation dominated stage at the GUT scale with $T \simeq 10^{15}\,\mathrm{GeV}$, with $m \ll T$ and $k_p   \simeq T$ corresponding to $E_k(t) \simeq T $ and $\gamma_k \gg 1$ (ultrarelativistic). With the number of ultrarelativistic degrees of freedom $g_\text{eff} \simeq 100$ the expansion  rate is
\be H(t) \simeq 1.66 \sqrt{g_\text{eff}}~ \frac{T^2(t)}{M_\text{Pl}} \simeq 10^{-2}\,T(t) \,,\label{Hrad}\ee and it follows that
\be   \frac{\omega^{'}_k(\eta)}{\omega^2_k(\eta)} \ll 1 \,.\label{vali}\ee This analysis clarifies the separation of scales: the Hubble expansion rate $H(t)\ll E_k(t)$, namely there are many oscillations of the field during a Hubble time and the ratio is further suppressed by large local Lorentz factors.
This ratio  becomes smaller as the scale factor grows and the Hubble rate slows, thereby improving the reliability of the adiabatic expansion. For example, today $H(t_0) \simeq H_0 \simeq 10^{-42}\,\mathrm{GeV}$, which is much smaller than the typical particle physics scales even for very light   axion-like (DM) candidates.

Therefore we adopt the   ratio
\be  \frac{H(t)}{E_k(t)} \ll 1 \,, \label{adpara}\ee as the small, dimensionless \emph{adiabatic} expansion parameter. The physical interpretation of this (small) ratio is clear: typical particle physics degrees of freedom feature wavelengths that are much smaller than the particle horizon proportional to the Hubble radius   at any given time (see discussion section below for caveats).

Consequently, when considering the term $a''/a$ in the equation of motion (\ref{EomFT}), we find that
\be \frac{a''}{a\,\omega^2_k} = 2 \Big( \frac{\dot{H}}{2\,E^2_k} + \frac{H^2}{E^2_k} \Big) = \alpha\,\frac{H^2}{E^2_k} ~~;~~ \alpha\simeq 0 ~~~ (RD)~;~ \alpha \simeq \frac{1}{2}~~~(MD)\,. \label{addot} \ee Therefore the ratio $a''/\omega^2_k a$ is of second adiabatic order and can be safely neglected to the leading adiabatic order which we will pursue in this study, justifying the simplification of the mode equations to (\ref{eqnofmot}).

In this article we consider the zeroth-adiabatic order with the mode functions given by
\be g_k (\eta) = \frac{e^{-i\,\int^{\eta}_{\eta_i}\,\omega_k(\eta')\,d\eta'}}{\sqrt{2\,\omega_k(\eta)}} \label{WKB0ord}\ee postponing to future study higher adiabatic corrections (see discussion section below). The phase of the mode function has an immediate interpretation in terms of comoving time and the local comoving energy (\ref{comoener}), namely
\be e^{-i\,\int^{\eta}_{\eta_i}\,\omega_k(\eta')\,d\eta'} = e^{-i\,\int^{t}_{t_i}\,E_k(t')\,dt'}\,,\label{phase}\ee which is a natural and straighforward generalization of the phase of \emph{positive frequency particle states in Minkowski space-time}.

\section{Particle interpretation: Adiabatic Hamiltonian}

Unlike in Minkowski space-time where the full Lorentz group unambiguously leads to a description of particle states associated with Fock states that transform irreducibly   and are characterized by mass and spin, the definition of particle states in an expanding cosmology without a global time-like Killing vector is more subtle\cite{birrell,fullbook,mukhabook,parkerbook,birford,parker}.

Our goal is to study particle decay implementing the adiabatic approximation described above, focusing on the leading, zeroth order contribution with the mode functions (\ref{WKB0ord}). Field quantization in terms of these modes entail that the creation and annihilation operators of the adiabatic particle states depend on time so that the quantum field obeys the (free field) Heisenberg equations of motion.  Passing to the interaction picture to obtain the transition amplitudes and probabilities, we would need the explicit time dependence of the creation and annihilation operators. In this section we show explicitly that to leading adiabatic order the operators that create and annihilate the adiabatic states are \emph{time independent}. This is an important simplification that allows the calculation of matrix elements in a straightforward manner.

In order to establish a clear identification of the zeroth order adiabatic modes with particles we analyze the free-field Hamiltonian, which in terms of the conformally rescaled field operators is given by
\begin{equation}
 H(\eta) =\frac12\int d^3x\,\{ \pi^2+(\nabla\chi)^2+\mathcal{M}^2(\eta)\chi^2\}\,.
\end{equation}
Writing the field operators in terms of their Fourier expansions, we have
\bea
  \chi(\vec x,\eta) & = & \frac{1}{\sqrt V}\sum_k[a_kg_k(\eta)e^{i\vec
      k\cdot\vec x} + a_k^{\dagger}g^*_k(\eta)e^{-i\vec k\cdot\vec x}],\label{chifi} \\
  \pi(\vec x,\eta) & = & \chi'(\vec x,\eta) = \frac{1}{\sqrt
    V}\sum_k[a_kg'_k(\eta)e^{i\vec k\cdot\vec x} +
    a_k^{\dagger}g_k^*{}'(\eta)e^{-i\vec k\cdot\vec x}]  \,.\label{chipi}
\eea

Integrating over $d^3x$, gathering terms and neglecting the term $a''/a$ in (\ref{EomFT}) as discussed above, we find
\begin{align}
    H(\eta) &= \frac{1}{2}\sum_k\Bigl{\{}
    a_k^\dagger a_k\bigl(|g_k'|^2+\omega^2_k(\eta)\,|g_k|^2\bigr)
    +a_k a_{-k}\bigl((g_k')^2+\omega^2_k(\eta)(g_k)^2\bigr) + h.c.
    \Bigr{\}}\\
    &\equiv\frac{1}{2}\sum_k\Bigl{\{}\Omega_k(\eta)a_k^\dagger a_k
    +\Delta_k(\eta)a_k a_{-k} + h.c.\,\Bigr{\}}.
\end{align}
We can now expand these coefficients $\Omega_k(\eta)$ and $\Delta_k(\eta)$ in terms of the functions $W_k(\eta)$ by using the explicit form of the mode functions
  \begin{equation}
  g_k(\eta) = \frac{\mode}{\sqrt{2W_k(\eta)}}\,;\, g'_k(\eta)  = -iW_k(\eta)g_k(\eta)\Big[1 -i\frac{W'_k(\eta)}{2W^2_k(\eta)}\Big]\label{ggprim}
  \end{equation}
and using  the relation (\ref{WKBsol}) the  frequencies $\Omega_k(\eta);\Delta_k(\eta)$ can be written as
\begin{equation}
    \Omega_k(\eta) = |g_k|^2\Bigl(2W_k^2+\frac{W_k''}{2W_k}-\frac{W_k'{}^2}{2W_k^2}\Bigr)
    ,\quad
    \Delta_k(\eta) = (g_k)^2\Bigl(\frac{W_k''}{2W_k}-\frac{W_k'{}^2}{2W_k^2}-iW'_k\Bigr).
\end{equation}
It is convenient to introduce
 \be \alpha_k(\eta)\equiv \frac{W_k''}{2W_k}-\frac{W_k'{}^2}{2W_k^2}\,,\label{alfa}\ee
  which allows us to rewrite the Hamiltonian as
\begin{equation}
    H(\eta) = \frac{1}{2}\sum_k
    \begin{pmatrix}
    a^\dagger_k & a_{-k}
    \end{pmatrix}
     \begin{pmatrix}
    |g_k|^2(\alpha_k+2W_k^2)  & (g_k^*)^2( \alpha_k+iW'_k)\\
    (g_k)^2(\alpha_k -iW'_k) & |g_k|^2(\alpha_k+2W_k^2)
    \end{pmatrix}
     \begin{pmatrix}
    a_k \\
    a^\dagger_{-k}
    \end{pmatrix}
\end{equation}
This Hamiltonian can be diagonalized by a time-dependent  Bogoliubov transformation. We do  this in two steps. First we write
\be \tilde{a}_k(\eta) =   a_k \,e^{-i\int^\eta W_k(\eta')\,d\eta'}\, e^{-i\theta_k(\eta)/2}\,,\label{atil}\ee and choose $\theta_k(\eta)$ to absorb  the phase of $\Delta_k$, i.e., $\tan\theta_k(\eta) = W'_k(\eta)/\alpha_k(\eta)$. Then
\begin{equation}
    H(\eta) = \frac{1}{2}\sum_k
    \begin{pmatrix}
    \tilde{a}^\dagger_k & \tilde{a}_{-k}
    \end{pmatrix}
     \begin{pmatrix}
    \Omega_k(\eta)  & \Dk(\eta)\\
    \Dk(\eta) & \Omega_k(\eta)
    \end{pmatrix}
     \begin{pmatrix}
    \tilde{a}_k \\
    \tilde{a}^\dagger_{-k}
    \end{pmatrix},
\end{equation}
where
\be  \Omega_k(\eta)   =   \frac{1}{2W_k}(\alpha_k(\eta)+2W_k^2(\eta) )~~;~~ \Dk  =
\frac{1}{2W_k}\sqrt{\alpha_k^2(\eta)+(W'_k(\eta))^2  }\,.\label{OmeDelk}\ee      We introduce the Bogoliubov transformation  to a new set of   creation and annihilation operators $\hat{b}^\dagger_{\vk}$, $\hat{b}_{\vk}$ as
\be\tilde{a}^\dagger_{\vk} = u_k(\eta) \,\hat{b}^\dagger_{\vk}+v_k \,(\eta)  \hat{b}_{-\vk}~~;~~ \tilde{a}_{\vk} = u_k(\eta) \, \hat{b}_{\vk}+v_k(\eta) \,\hat{b}^\dagger_{-\vk}\,,\label{bogo}\ee noting that  $u_k,v_k$ are real  functions of $\eta$ and  $|\vk|$ only.    For the $\hat{b}_{\vk}$, $\hat{b}^\dagger_{\vk}$ to obey the canonical commutation relations, it follows that  $u_k^2-v^2_k = 1$. Then the Hamiltonian can be written
\begin{align}
    H(\eta) &= \frac12\sum_k
    \begin{pmatrix}
    \hat{b}^\dagger_{\vk} & \hat{b}_{-\vk}
    \end{pmatrix}
     \begin{pmatrix}
    u_k  & v_k\\
    v_k & u_k
    \end{pmatrix}
     \begin{pmatrix}
    \Omega_k  & \Dk\\
    \Dk & \Omega_k
    \end{pmatrix}
    \begin{pmatrix}
    u_k  & v_k\\
    v_k & u_k
    \end{pmatrix}
     \begin{pmatrix}
    \hat{b}_{\vk} \\
    \hat{b}^\dagger_{-\vk}
    \end{pmatrix}\\
    &=
    \frac12\sum_k
    \begin{pmatrix}
    \hat{b}^\dagger_{\vk} & \hat{b}_{-\vk}
    \end{pmatrix}
     \begin{pmatrix}
    (u^2_k + v^2_k)\Omega_k +2u_kv_k\Dk
    & (u^2_k + v^2_k)\Dk +2u_kv_k\Omega_k \\
    (u^2_k + v^2_k)\Dk +2u_kv_k\Omega_k
    & (u^2_k + v^2_k)\Omega_k +2u_kv_k\Dk
    \end{pmatrix}
     \begin{pmatrix}
    \hat{b}_{\vk} \\
    \hat{b}^\dagger_{-\vk}
    \end{pmatrix},
\end{align}
and the $u_k$ and $v_k$ chosen to make off-diagonal terms vanish. Then writing $u_k = \cosh{\phi_k}$ and $v_k = \sinh{\phi_k}$, we find
\begin{equation}
    \tanh{2\phi_k} = -\frac{\Dk}{\Omega_k}.
\end{equation}

In the second step we absorb the fast phases into the redefinition
\be \hat{b}_{\vk}\, = e^{-i\int^\eta W_k(\eta')\,d\eta'}\, b_{\vk}~~;~~ \hat{b}^\dagger_{\vk}\, = e^{i\int^\eta W_k(\eta')\,d\eta'}\, b^\dagger_{\vk} \,,\label{secdef} \ee
in terms of which the  Hamiltonian can be written as
\be
H(\eta) =\sum_k \omega_k(\eta) \Bigl( {b}_{\vk}^\dagger(\eta)_{\vk}\,  {b}_{\vk}(\eta) + \tfrac12\Bigr)\,. \label{Hdiag}
\ee
This is a remarkable result:   the new operators $b^\dagger_{\vk}, b_{\vk}$ define a Fock Hilbert space of \emph{adiabatic eigenstates}, the exact  frequencies of which are the \emph{zeroth order adiabatic frequencies} $\omega_k(\eta) = \sqrt{k^2 + m^2\,a^2(\eta)}$. We emphasize that $b^\dagger_{\vk}(\eta),b_{\vk}(\eta)$ depend explicitly on time because the Bogoliubov coefficients $u_k(\eta),v_k(\eta)$ depend on time, while the original operators $a_{\vk},a^\dagger_{\vk}$ are time independent in the Heisenberg picture. This is also clear by inverting the relations (\ref{bogo}), and using (\ref{atil}) the redefinition (\ref{secdef}) along with $u^2_k-v^2_k=1$,  we find
\bea b^\dagger_{\vk}(\eta) & = &  u_k(\eta) \,e^{-i\theta_k(\eta)/2}\,a^\dagger_{\vk} + v_k(\eta) \,e^{ i\theta_k(\eta)/2}\,e^{-2i\int^\eta W_k(\eta')\,d\eta'}\,a_{-\vk} \label{bdaginv}\\
b_{\vk}(\eta) & = &  u_k(\eta) \,e^{i\theta_k(\eta)/2}\,a_{\vk} + v_k(\eta) \,e^{- i\theta_k(\eta)/2}\,e^{2i\int^\eta W_k(\eta')\,d\eta'}\,a^\dagger_{-\vk}\,. \label{binv}\eea

Using (\ref{WKBsol}) and the adiabatic expansion (\ref{adexp}) it is straightforward to find that
\be u_k(\eta) = 1 + \mathcal{O}\Big((\omega'_k(\eta))^2, \omega^{''}_k(\eta)\Big)~~;~~ v_k(\eta)\simeq \mathcal{O}\Big((\omega'_k(\eta))^2, \omega^{''}_k(\eta)\Big)\,.\label{uvad}\ee
Hence, \emph{to zeroth order in the adiabatic expansion} $b_{\vk} = a_{\vk}$ and the annihilation and creation operators of \emph{adiabatic particle} states are independent of time. Time dependence of the operators $b_{\vk},b^\dagger_{\vk}$ emerges at \emph{second order} in the adiabatic expansion.

Therefore, the study in this section justifies our identification of particle states to leading (zeroth) order in the adiabatic expansion, namely the \emph{time independent}  operators  $a^\dagger,a$ create and annihilate  zeroth order adiabatic  particle states of time dependent frequency $\omega_k(\eta)$. This is important because below we cast the interaction picture in terms of these operators and the mode functions $g_k(\eta)$. The analysis above explicitly shows the consistency of this approach to leading order in the adiabatic approximation. In higher order the time evolution of the operators $b,b^\dagger$ entail particle production\cite{parker,birrell,fullbook,mukhabook,parkerbook,birford,dunne,wini}, an important aspect that will be relegated to future study (see discussion section below). In the analysis that follows we will consider the leading (zeroth) order adiabatic modes.

\section{The Interaction picture in Cosmology}\label{subsec:IPcosmo}
The creation and annihilation operators $a_{\vk},a^\dagger_{\vk}$  for each respective field    define Fock states,  with a vacuum state $|0\rangle$ defined by $a_{\vk}\,|0\rangle=0 $. Since    to leading order  in the adiabatic approximation $a,a^\dagger$ coincide with $b,b^\dagger$  associated with single particle adiabatic states, it follows that  $a^\dagger_{\vk} \,|0\rangle$  are identified (to this order) with the single particle states associated with the mode functions(\ref{WKB0ord}).

In the Schr\"odinger picture, quantum states obey
\begin{equation}
i\frac{d}{d\eta}\ket{\Psi(\eta)} = H(\eta)\ket{\Psi(\eta)}\,,\label{sch}
\end{equation}
where in general the Hamiltonian carries explicit $\eta$ dependence.  The solution of (\ref{sch}) is given in terms of the unitary time evolution operator $U(\eta,\eta_0)$, namely
$ \ket{\Psi(\eta)} = U(\eta,\eta_0)\ket{\Psi(\eta_0)}$,  $U(\eta,\eta_0)$ obeys
\begin{equation}
i\frac{d}{d\eta}U(\eta,\eta_0) = H(\eta)U(\eta,\eta_0)~~;~~ U(\eta_0,\eta_0) =1 \,.\label{Uofeta}
\end{equation}

Consider a Hamiltonian that can be written as $H(\eta) = H_0(\eta)+H_i(\eta)$, where $H_0(\eta)$ is the free theory Hamiltonian and $H_i(\eta)$ the interaction Hamiltonian. In the absence of interactions with $H_i=0$, the time evolution operator of the free theory $U_0(\eta,\eta_0)$ obeys
\begin{equation}
i\frac{d}{d\eta}U_0(\eta,\eta_0) = H_0(\eta)U_0(\eta,\eta_0),
\quad
-i\frac{d}{d\eta}U^{-1}_0(\eta,\eta_0) = U^{-1}_0(\eta,\eta_0)H_0(\eta),
\quad U(\eta_0,\eta_0)=1 \,. \label{freeH0}
\end{equation}

 It is convenient to pass to the interaction picture, where the operators evolve with the free field Hamiltonian and the states carry the time dependence from the interaction, namely
\be |\Psi(\eta)\rangle_I = U^{-1}_0(\eta,\eta_0)\,|\Psi(\eta)\rangle\,, \label{psiip}\ee and their time evolution is given by
\be |\Psi(\eta)\rangle_I = U_I(\eta,\eta_0) \,|\Psi(\eta_0)\rangle_I ~~;~~ U_I(\eta,\eta_0) = U^{-1}_0(\eta,\eta_0)\,U(\eta,\eta_0)\,.  \label{evolip}\ee The unitary time evolution operator in the interaction picture $U_I(\eta,\eta_0)$ obeys
\begin{equation}
i\frac{d}{d\eta}U_I(\eta,\eta_0) = H_I(\eta)U_I(\eta,\eta_0) \quad H_I(\eta) = U^{-1}_0(\eta,\eta_0)H_i(\eta)U_0(\eta,\eta_0)~~;~~ U_I(\eta_0,\eta_0) =1\,.  \label{Uip}
\end{equation} For the conformal action (\ref{conformalaction}) it follows that
\be H_I(\eta) = \lambda \, a(\eta) \int d^3x ~ \chi_1(\vx,\eta)\,:\chi^2_2(\vx,\eta):\,, \label{HI}\ee where the fields are given by the free field expansion (\ref{quant}) with the mode functions solutions of (\ref{geqnofmot},\ref{wron}) and time independent creation and annihilation operators for the respective fields. The perturbative solution of eqn. (\ref{Uip}) is
\be U_I(\eta,\eta_0) = 1-i\int^{\eta}_{\eta_0} H_I(\eta_1)\,d\eta_1 + (-i)^2 \int^{\eta}_{\eta_0}  \int^{\eta_1}_{\eta_0} H_I(\eta_1)\,H_I(\eta_2)\,d\eta_1 \,d\eta_2 +\cdots \label{Uippert}\ee

\vspace{1mm}

\textbf{Amplitudes and probabilities in perturbation theory.}

Before we consider the \emph{non-perturbative} Wigner-Weisskopf method, we study the transition amplitudes and probabilities in perturbation theory as this will yield a clear interpretation of the results.

Let us consider the amplitude for the decay process $\chi_1 \rightarrow 2\,\chi_2$ given by
\be \mathcal{A}_{1\rightarrow 2 2}(\eta,\eta_i) = \langle 1^{(2)}_{\vp}, 1^{(2)}_{\vq}|\,U_I(\eta,\eta_i)\,|1^{(1)}_{\vk}\rangle \,,\label{amp} \ee where $|1^{(a)}_{\vec{p}}\rangle, a=1,2$ are the single particle states associated with the respective fields. With the expansion (\ref{Uippert}) we find to lowest order in perturbation theory,
\bea  \mathcal{A}_{1\rightarrow 2 2}(\eta,\eta_i)  & =  & -i\lambda \int^{\eta}_{\eta_i} d\eta' \, a(\eta')\,\int d^3 x \,\langle 1^{(2)}_{\vp}, 1^{(2)}_{\vq}|\chi_1(\vx,\eta')\,\chi^2_2(\vx,\eta')\,|1^{(1)}_{\vk}\rangle \nonumber \\
& = & -2i\frac{\lambda}{V^{1/2}}\,\int^{\eta}_{\eta_i} d\eta' \, a(\eta')\,g^{(1)}_k(\eta')\,(g^{(2)}_p(\eta'))^*\,(g^{(2)}_q(\eta'))^*\,\delta_{\vk,\vp+\vq} \,. \label{amp12}\eea  The total transition probability is given by
\be \mathcal{P}_{1\rightarrow 2 2}(\eta,\eta_i) = \frac{1}{2 !} \sum_{\vp}\sum_{\vq} |\mathcal{A}_{1\rightarrow 2 2}(\eta,\eta_i)|^2\,, \label{totprob}\ee and taking the continuum limit yields
\be \mathcal{P}_{1\rightarrow 2 2}(\eta,\eta_i) = \int^{\eta}_{\eta_i} d\eta_2 \, \int^{\eta}_{\eta_i} d\eta_1\,\Sigma_k(\eta_2;\eta_1)\,, \label{pofsig}\ee where
\be \Sigma_k(\eta;\eta') = 2\lambda^2 a(\eta)\,a(\eta')\,(g^{(1)}_k(\eta))^*\,g^{(1)}_k(\eta')\,\int \frac{d^3p}{(2\pi)^3}~ g^{(2)}_p(\eta)\,g^{(2)}_q(\eta)\,(g^{(2)}_k(\eta'))^*\,(g^{(2)}_q(\eta'))^* ~~;~~ q=|\vk-\vp| \,.\label{sigma}\ee Noting the property
\be (\Sigma_k(\eta;\eta'))^* = \Sigma_k(\eta';\eta)\,, \label{prop}\ee and introducing the identity $\Theta(\eta_2-\eta_1) + \Theta(\eta_1-\eta_2) =1$, relabelling terms and using the property (\ref{prop}), we find
\be \mathcal{P}_{1\rightarrow 2 2}(\eta,\eta_i) = 2 \int^{\eta}_{\eta_i} d\eta_2 \int^{\eta_2}_{\eta_i} d\eta_1 \,\mathrm{Re}[\Sigma_k(\eta_2;\eta_1)]\,.  \label{probfin}\ee We  define  the \emph{transition rate}
\be \Gamma(\eta) \equiv \frac{d}{d\eta}\mathcal{P}_{1\rightarrow 2 2}(\eta,\eta_i)  = 2 \int^{\eta}_{\eta_i} d\eta_1\,\mathrm{Re}[\Sigma_k(\eta;\eta_1)] \,.\label{defgamma}\ee We emphasize to the reader that in typical S-matrix calculations in Minkowski spacetime, the presence of a time-like Killing vector (and the implementation of the infinite time limit) lead to a \emph{time independent transition rate} and subsequently a standard exponential decay law. In FRW spacetime, this approach is in general invalid.  Rather, the transition rate introduced above will define the decay law obtained within the non-perturbative Wigner-Weisskopf framework described below.

\section{Wigner--Weisskopf theory in cosmology}\label{sec:WW}

The quantum field theoretical Wigner-Weisskopf method has been introduced in refs.\cite{boyww,boycosww}, where the reader is referred to for more details. As discussed in these references, this method is manifestly unitary and leads to a non-perturbative systematic description of transition amplitudes and probabilities directly in real time. Here we describe the main aspects of its implementation within the cosmological setting.  Consider an  interaction picture state $\ket{\Psi(\eta)}_I = \sum_n C_n(\eta)\ket{n}$, expanded in the Hilbert space of the free theory; these are the Fock states associated with the annihilation and creation operators $a_{\vk},a^\dagger_{\vk}$ of the free field expansion (\ref{chifi}) for each field. Inserting into \eqref{Uip} yields an \emph{exact}  set of coupled  equations for the coefficients
\begin{equation}
i\frac{d}{d\eta} C_n(\eta) = \sum_m C_m(\eta)\bra{n}H_I(\eta)\ket{m}.
\end{equation}

\begin{figure}[ht!]
\begin{center}
\includegraphics[height=3in,width=3in,keepaspectratio=true]{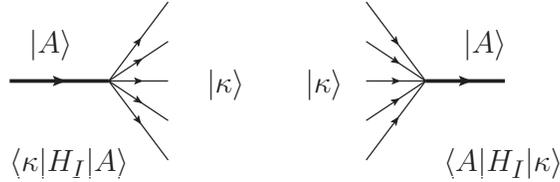}
\caption{Transitions $|A\rangle \leftrightarrow |\kappa\rangle$ in first order in $H_I$.}
\label{fig1:coupling}
\end{center}
\end{figure}

In principle this is an infinite hierarchy of integro-differential equations for the coefficients $C_n(\eta)$; progress can be made, however, by considering states connected
by the interaction Hamiltonian to a given order in the interaction. Consider that initially the state is $\ket{A}$ so that $C_A(\eta_i) =1\,;\,C_{\kappa}(\eta_i) =0$ for $\ket{\kappa}\neq \ket{A}$, and consider a  first order transition  process
$\ket{A}\rightarrow\ket{\kappa}$  to    intermediate multiparticle states $\ket{\kappa}$ with transition matrix elements $\langle \kappa|H_I(\eta)|A\rangle$. Obviously the state $\ket{\kappa}$ will be connected to other multiparticle states $\ket{\kappa'}$  different from $\ket{A}$ via $H_I(\eta)$. Hence for example up to second order in the interaction, the  state $\ket{A}\rightarrow \ket{\kappa}\rightarrow \ket{\kappa'}$. Restricting the hierarchy to \emph{first order transitions} from the initial state $\ket{A} \leftrightarrow \ket{\kappa}$ results in a coupled set of equations
\bea
i\frac{d}{d\eta} C_A(\eta) &= &  \sum_\kappa C_\kappa(\eta)\bra{A}H_I(\eta)\ket{\kappa}\label{eqA}\\
i\frac{d}{d\eta} C_\kappa(\eta) &= & C_A(\eta)\bra{\kappa}H_I(\eta)\ket{A}~~;~~ C_A(\eta_i) =1\,;\,C_{\kappa}(\eta_i) =0 \,. \label{eqK}
\eea These processes are depicted in fig. (\ref{fig1:coupling}).

 Equation (\ref{eqK}) with $C_{\kappa}(\eta_i)=0$ is formally solved by
 \be C_{\kappa}(\eta) = -i\int^{\eta}_{\eta_i}\,\bra{\kappa}H_I(\eta')\ket{A}\,C_{A}(\eta')\,d\eta' \,, \label{kappapop}\ee
 and inserting this  solution into equation (\ref{eqA}) we find
\begin{equation}\label{diffeqCA}
\frac{d}{d\eta} C_A(\eta) = -\int_{\eta_i}^\eta d\eta' \,
\Sigma_A(\eta,\eta')~ C_A(\eta')\,,
\end{equation} where we have introduced   the \emph{self-energy}
\be \Sigma_A(\eta;\eta') =
\sum_\kappa \bra{A}H_I(\eta)\ket{\kappa}
\bra{\kappa}H_I(\eta')\ket{A} \,. \label{sigmaA}\ee  This integro-differential equation  with \emph{memory} yields a non-perturbative solution for the time evolution of the amplitudes and probabilities. In Minkowski space-time and in frequency space, this is recognized as a Dyson resummation of self-energy diagrams, which upon Fourier transforming back to real time, yields the usual exponential decay law\cite{boyww,boycosww}. Introducing the solution for $C_A(\eta)$ back into (\ref{eqK}) yields the build-up of the \emph{population} of ``daughter'' particles.

The equation (\ref{diffeqCA}) is in general very difficult to solve, but progress can be made under the weak coupling assumption by invoking the \emph{Markovian} approximation. A systematic implementation of this approximation begins by introducing
\be
\mathcal{E}_A(\eta,\eta')  \equiv \int_{\eta_i}^{\eta'} \Sigma_A(\eta,\eta'')\,d\eta'' \,, \label{Evar}
\ee
such that
\be \frac{d}{d\eta'}\,\mathcal{E}_A(\eta,\eta')=\Sigma_A(\eta,\eta') \,,  \label{trick}\ee with the condition
\be   \mathcal{E}_A(\eta,\eta_i)=0 \,. \label{condiEA}\ee

Then \eqref{diffeqCA} can be written as
\begin{equation}
\frac{d}{d\eta} C_A(\eta)  =  -\int_{\eta_i}^\eta d\eta' \,
\frac{d}{d\eta'}\mathcal{E}_A(\eta,\eta')\,  C_A(\eta')
\end{equation}
which can be integrated by parts to yield
\begin{equation}
\frac{d}{d\eta} C_A(\eta) = -\mathcal{E}_A(\eta,\eta) C_A(\eta) +\int_{\eta_i}^\eta d\eta' \,
\mathcal{E}_A(\eta,\eta') \frac{d}{d\eta'}C_A(\eta').
\end{equation}
Since $\mathcal{E}_A \propto \mathcal{O}(H^2_I)$ the first term on the right hand side is of order $H^2_I$, whereas the second is $\order{H^4_I}$ because $dC_A(\eta)/d\eta  \propto \mathcal{O}(H^2_I)$. Therefore  to leading order in the interaction, the evolution equation for the amplitude becomes
\begin{equation}
\frac{d}{d\eta} C_A(\eta) = -\mathcal{E}_A(\eta,\eta) C_A(\eta)  \,,\label{LOeq}
\end{equation} with   solution
\begin{equation}
C_A(\eta) = \exp\Bigl(-\int^\eta_{\eta_i}\mathcal{E}_A(\eta',\eta')\,d\eta'\Bigr)\,C_A(\eta_i)\,. \label{CAfina}
\end{equation} This expression clearly highlights the non-perturbative nature of the Wigner-Weisskopf approximation. The imaginary part of the self energy $\Sigma_A$ yields a \emph{renormalization} of the frequencies which we will not pursue here\cite{boycosww,boyww}, whereas the real part gives the decay rate, with
\be |C_A(\eta)|^2 = e^{-\int^\eta_{\eta_i} \Gamma_A(\eta') d\eta'} \,|C_A(\eta_i)|^2 ~~;~~ \Gamma_A(\eta) = 2 \int^\eta_{\eta_i}d\eta_1\,\mathrm{Re}\,[\Sigma_A(\eta,\eta_1)]   \,. \label{probbaA}\ee

Finally, the amplitude for the decay product state $\ket{\kappa}$ is obtained by inserting the amplitude (\ref{CAfina})  into (\ref{kappapop}), and the \emph{population} of the daughter particles is $|C_{\kappa}(\eta)|^2$.

In our study the state $\ket{A}$ is a single particle state of momentum $\vk$ of the decaying parent particle.

\subsection{Disconnected vacuum diagrams}\label{subsec:disc}
\begin{figure}[ht!]
\begin{center}
\includegraphics[height=3.5in,width=4.5in,keepaspectratio=true]{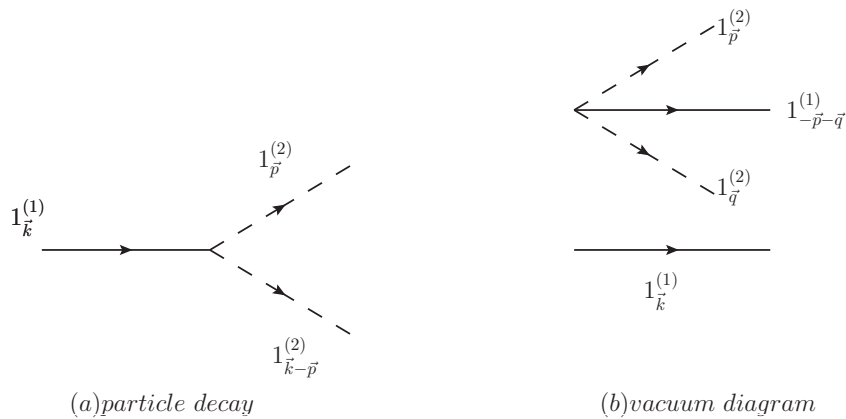}
\caption{Decay and vacuum diagrams for  $\ket{A}=\ket{1_k^{(1)}}$ to first order in $H_I$. Solid lines single particle states of the field $\chi_1$, dashed lines are single particle states of the field $\chi_2$.}
\label{fig2:decayvacuum}
\end{center}
\end{figure}

Before we set out to obtain the self-energy and decay law for a single particle state of the field $\chi_1$ into two particles of the field $\chi_2$ we must clarify the nature of the vacuum diagrams. Consider the initial single particle state $\ket{A}=\ket{1_{\vk}^{(1)}}$  and the set of intermediate states connected to this state by the interaction Hamiltonian to first order. There are \emph{two different} contributions: a): the decay process  $\ket{1_{\vk}^{(1)}} \rightarrow \ket{1_{\vp}^{(2)};1_{\vk-\vp}^{(2)}}$ in which the initial state is annihilated and the two particle state produced, and b): a  \emph{four particle state} in which the initial state evolves unperturbed and a \emph{three particle state} $\ket{1_{\vp}^{(2)};1_{\vq}^{(2)};1_{-\vp-\vq}^{(1)}}$ is created out of the vacuum by the perturbation. These contributions  are   depicted in fig. (\ref{fig2:decayvacuum}).

\begin{figure}[ht!]
\begin{center}
\includegraphics[height=3in,width=4in,keepaspectratio=true]{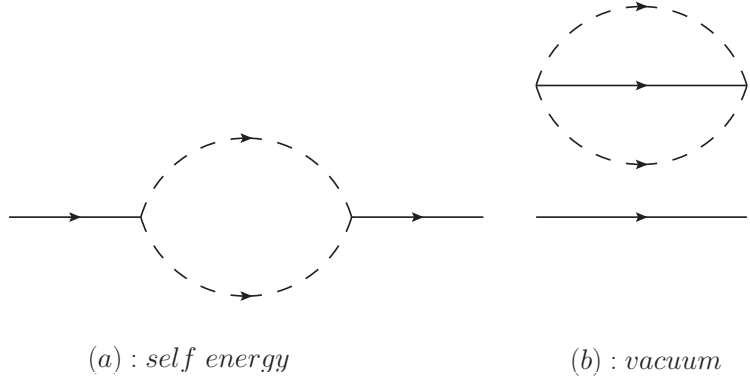}
\caption{Contributions to the self-energy   for decay (a) and vacuum diagram (b) for  $\ket{A}=\ket{1_k^{(1)}}$ to first order in $H_I$ with the same notation as in fig.(\ref{fig2:decayvacuum}). }
\label{fig3:decayvacuumse}
\end{center}
\end{figure}

These processes yield two different contributions to $\sum_\kappa \bra{1_{\vk}^{(1)}}H_I(\eta)\ket{\kappa}
\bra{\kappa}H_I(\eta')\ket{1_{\vk}^{(1)}} $, depicted in fig. (\ref{fig3:decayvacuumse}).

The second disconnected diagram (b) corresponds to the ``dressing'' of the vacuum. This is clearly understood by considering the initial state to be the non-interacting vacuum state $\ket{0}$; it is straightforward to repeat the analysis above to obtain the closed set of leading order equations that
describe the build-up of the full interacting vacuum state. One finds that diagram (b) without the non-interacting single particle state precisely describes the ``dressing'' of the vacuum state.
Clearly, similar disconnected diagrams enter the evolution of \emph{any} initial state. In the case under consideration, namely the decay of single particle states, the disconnected diagram (b) \emph{does not} contribute to the decay but to the definition of a single particle state obtained out of the full vacuum state.
In   S-matrix theory these disconnected diagrams are cancelled by dividing \emph{all transition matrix elements} by $\bra{0}S\ket{0}$.  Within the Wigner-Weisskopf framework, we write the amplitude for the single particle state $\ket{A} = \ket{1^{(1)}_{\vk}}$ as
\be C_A(\eta) = \tilde{C}_A(\eta)\,\tilde{C}_0(\eta)\, \label{fact}\ee where $\widetilde{C}_0(\eta)$ is the amplitude for the interacting vacuum state obeying
\be \frac{d}{d\eta} \tilde{C}_0(\eta) = -\mathcal{E}_0(\eta,\eta)\,\tilde{C}_0(\eta) \,, \label{vacamp}\ee where
\be \mathcal{E}_0(\eta,\eta')  \equiv \int_{\eta_i}^{\eta'} \Sigma^{(b)}_A(\eta,\eta'')\,d\eta'' \,, \label{vacSE}  \ee and $\Sigma^{(b)}_A(\eta,\eta'')$ is the vacuum self-energy diagram $(b)$ in figure (\ref{fig3:decayvacuumse}). With the total self energy given by the sum of the decay $(a)$ and vacuum ($b)$ diagrams as in figure (\ref{fig3:decayvacuumse}), it follows that the  amplitude $\tilde{C}_A(\eta)$ obeys
 \begin{equation}
\frac{d}{d\eta} \tilde{C}_A(\eta) = -\mathcal{E}^{(a)}_A(\eta,\eta) \tilde{C}_A(\eta)\,,\label{tilLOeq}
\end{equation}   where $\mathcal{E}^{(a)}_A$ is determined \emph{only} by the connected (decay) self energy diagram $(a)$. This is precisely the same as dividing by the vacuum matrix element in  S-matrix theory where similar diagrams arise in Minkowski space time with a similar interpretation\cite{boycosww,boyww}. This is tantamount to redefining the single particle states as built from the full vacuum state. Therefore  we neglect diagram (b).  We emphasize that this is in contrast with the method proposed in ref.\cite{vilja} wherein following ref.\cite{spangdecay} the disconnected diagram (b) is kept in the calculation of the decay process.

Now we are able to calculate the general form of the decay law   by considering the decay process $\chi_1\rightarrow2\chi_2$ in the interacting theory with $H_I(\eta)$ given by (\ref{HI}) to leading order in $\lambda$ and keeping only the connected diagrams.

  The initial state corresponds to   single particle state of the $\chi_1$ field
    $\ket{ A} = \ket{1_k^{(1)}}$,  and the decay process corresponds to a transition to    the state $|\kappa\rangle = \ket{1_{\vp}^{(2)};1_{\vq}^{(2)}}$.
  Then
  \bea
  \bra{1_{\vp}^{(2)};1_{\vq}^{(2)}}H_I(\eta')\ket{1_k^{(1)}} &=&\frac{2\lambda \,a(\eta')}{V^{1/2}}
     g_k^{(1)}(\eta')g_p^{(2)}{}^*(\eta')g_q^{(2)}{}^*(\eta') \,\delta_{\vk,\vp+\vq},\,\nonumber \\
  \bra{1_k^{(1)}}H_I(\eta)\ket{1_{\vp}^{(2)};1_{\vq}^{(2)}} &= & \frac{2\lambda \, a(\eta)}{V^{1/2}}
    g_k^{(1)}{}^*(\eta)g_p^{(2)}(\eta)g_q^{(2)}(\eta)\,\delta_{\vk,\vp+\vq} \,. \label{mtxele}
  \eea Taking the continuum limit, summing over intermediate states, and accounting for the Bose symmetry factor in the final states  yields
  \bea && \Sigma_k(\eta,\eta') =  \frac{1}{2!} \sum_{\vp,\vq} \,  \bra{1_k^{(1)}}H_I(\eta)\ket{1_{\vp}^{(2)};1_{\vq}^{(2)}} \bra{1_{\vp}^{(2)};1_{\vq}^{(2)}}H_I(\eta')\ket{1_k^{(1)}}     \nonumber \\  & =  &
\frac{4\lambda^2}{2!}a(\eta)a(\eta')
   g_k^{(1)}(\eta')\,(g_k^{(1)}(\eta))^* \!\!\int\!\!\,\frac{d^3p}{(2\pi)^3}\,g_p^{(2)}(\eta)\,g^{(2)}_{|\vk-\vp|} (\eta)\,
   \,(g_p^{(2)}(\eta'))^* \,(g^{(2)}_{|\vk-\vp|}(\eta'))^*   \,. \label{siger} \eea

  This is \emph{precisely} the self-energy (\ref{sigma}) obtained in the perturbative description of the transition probability and amplitude, equation (\ref{pofsig}), which enters in the evolution equation (\ref{diffeqCA}) for the single (parent) particle.  Following the steps of the Markovian approximation leading up to the final result (\ref{probbaA}), we find
  \be |C_A(\eta)|^2 = |C_A(\eta_i)|^2\, \exp\biggl(-\int^\eta_{\eta_i} \,\Gamma_k(\eta') d\eta'\biggr)~~;~~ \Gamma_k(\eta') = 2\int^{\eta'}_{\eta_i}d\eta''\,\text{Re}\,\Sigma_k(\eta',\eta'')\,. \label{probaA}\ee This expression for the probability makes manifest the \emph{non-perturbative} nature of the Wigner-Weisskopf method.

\section{Decay law in leading adiabatic order.}

In this article we   study  the decay law in the theory described above to leading adiabatic order, namely zeroth order. The study of higher adiabatic order effects, primarily associated with the   production of particles by the cosmological expansion, is relegated to a future article (see discussion section below).

In the leading (zeroth)  order adiabatic approximation the mode functions are given by
\begin{equation}
g_{k}(\eta) = \frac{e^{-i\int_{\eta_i}^{\eta} \omega_k(\eta') d\eta'}}{\sqrt{2 \omega_k(\eta)}},
\quad \omega_k (\eta') = \sqrt{k^2 + m^2 a^2(\eta')}\,, \label{zeromo}
\end{equation}
 and $\Sigma_k$ takes the following form
\be
\Sigma_k(\eta,\eta')=\frac{2\lambda^2\, a(\eta)a(\eta')\,  }{\sqrt{2\omega_k^{(1)}(\eta)2\omega_k^{(1)}(\eta')}}  ~~\int\frac{d^3p}{(2\pi)^3}\frac{e^{i\int_{\eta'}^{\eta} \big[\omega^{(1)}_k(\eta'')-\omega_p^{(2)}(\eta'')-\omega_q^{(2)}(\eta'')\big] d\eta''}}{\sqrt{2\omega_p^{(2)}(\eta)2\omega_p^{(2)}(\eta')2\omega_q^{(2)}(\eta)2\omega_q^{(2)}(\eta')}}\,, \label{sigadzero}
\ee
where $q = | \vec{k} - \vec{p} |$. Obviously even to this order both the time and momentum integrals are daunting. However, progress is made by first considering the case of a massive parent particle decaying into two massless daughter particles. This study will reveal a path forward to the more general case of all massive particles.

\subsection{Massive parent, massless daughters in RD cosmology:}\label{subsec:massless}
Setting $m_2=0$,  the self energy becomes
\begin{equation}
\Sigma_k(\eta,\eta')=\frac{2\lambda^2\, a(\eta)a(\eta')\, e^{i\int_{\eta'}^{\eta} \omega_k(\eta'') d\eta''} }{\sqrt{2\omega_k^{(1)}(\eta)2\omega_k^{(1)}(\eta')}}  ~~\int\!\!\frac{d^3p}{(2\pi)^3}\frac{e^{-i(p+q)(\eta-\eta')}}{2p \, 2q} ~~;~~ q=|\vk-\vp|\,.\label{sigcos}
\end{equation}

The momentum integral in (\ref{sigcos}) is carried out \emph{exactly} by  introducing  a convergence factor after which it becomes
\be
I = \frac{1}{16 \pi^2} \!\!\int_0^{\infty} \frac{p^2\, dp}{p} \!\! \int_{-1}^{1} \frac{d(\cos(\theta))}{q} \, e^{-i(p+q)(s-i\epsilon)}, \, \epsilon \rightarrow 0^+, \quad s \equiv \eta - \eta'
\ee
Changing integration variables from $d(\cos(\theta))$ to $q  =|\vk-\vp|$ this integral becomes
 \be
I = \frac{1}{16 \pi^2 k} \!\int_0^{\infty} dp \, e^{-ip(s-i\epsilon)} \ \!\! \int_{|k-p|}^{|k+p|} dq \, e^{-iq(s-i\epsilon)}~~   = \frac{-ie^{-ik\,(\eta - \eta')}}{16 \pi^2 (\eta - \eta'-i\epsilon)}~~;~~ \epsilon \rightarrow 0^+ \label{integI} \,,
\ee  yielding
 \be
 \Sigma_k(\eta,\eta')  =\frac{\lambda^2\, a(\eta)\,a(\eta')\,\, e^{i\int_{\eta'}^{\eta} \omega_k(\eta'')\, d\eta''}\,e^{-ik(\eta-\eta')} }{16 \pi^2~\sqrt{\omega_k^{(1)}(\eta)\,\omega_k^{(1)}(\eta')}}~~\Bigg[-iP \Big(\frac{1}{\eta -\eta'} \Big)+\pi\delta(\eta-\eta')\Bigg],
\ee
where the Sokhotski-Plemelj theorem has been used in the last line. This expression is similar to that obtained in appendix (\ref{app:Mink}) in Minkowski space-time, where the scale factor is set to one and the frequencies are time independent (see eqn.~(\ref{sigminkfin})). The explicit time dependence obtained in Minkowski space-time in appendix~(\ref{app:Mink}) \emph{cannot} be gleaned in the usual calculations of decay rates via S-matrix theory where the initial and final times are taken to $\mp \infty$, respectively.

The decay width $\Gamma_k(\eta)$ is obtained from eqn.~(\ref{probaA}), and is given by
\be \Gamma_k(\eta) = \frac{\lambda^2\,a^2(\eta)}{8\pi \,\omega^{(1)}_k(\eta)}\,\,\frac{1}{2}\Big[1+ \mathcal{S}(\eta)\Big]\,, \label{Gamav1}\ee where  a factor of $\frac{1}{2}$ originates from the integration of the delta function in $\eta$ (the ``prompt'' term), and
\be \mathcal{S}(\eta) = \frac{2}{\pi}\, \int^{\eta}_{0} P[\eta,\eta'] \,\, \frac{\sin\big[A(\eta,\eta')\big]}{\eta-\eta'}\, d\eta'\,, \label{Sfunc}     \ee where we set $\eta_i=0$ and introduce
\bea P[\eta,\eta'] & = &  \frac{a(\eta')}{a(\eta)}\,\Bigg[\frac{\omega^{(1)}_k(\eta)}{\omega^{(1)}_k(\eta')}\Bigg]^{1/2} \,, \label{Pfac}\\ A(\eta,\eta') & = &  \int_{\eta'}^{\eta} \omega_k(\eta'') \, d\eta'' -k(\eta-\eta') \,. \label{Afac} \eea The expression for $\mathcal{S}$ can be simplified substantially, revealing a hierarchy of time scales associated with the adiabatic expansion in radiation domination, during which
\be a(\eta) = H_R\,\eta ~~;~~ H_R = H_0\,\sqrt{\Omega_R}\,.  \label{ard}\ee First we address the integral
\be J_k[\eta,\eta'] = \int^{\eta}_{\eta'}\,\omega^{(1)}_k(\eta'')\,d\eta'' = \int^{\eta}_{\eta'}\,\sqrt{k^2+m^2_1\,a^2(\eta'')}\,d\eta'' \,.  \label{intIk}\ee  To begin with we introduce the dimensionless variable (in what follows we suppress the $\eta$ dependence of $z$ to simplify notation)
\be  z= \omega_k(\eta)\,\eta = E_k(t) \,a(\eta) \, \eta = \frac{E_k(t)}{H(t)} \gg 1 \label{zvar}\ee
where $E_k(t) = \sqrt{k^2_{p}(t)+m^2}$ is the physical energy measured locally by a comoving observer with $k_p(t) = k/a(\eta)$   the physical momentum, and $H(t) = a'(\eta)/a^2(\eta) = 1/(\eta \,a(\eta))  $ during radiation domination, while $H(t) = 2/(\eta\,a(\eta))$ during matter domination. The dimensionless ratio (\ref{zvar}) is the \emph{inverse} of the adiabatic ratio $H(t)/E_k(t)$ (we have suppressed the momentum and $\eta$ dependence in z to simplify notation). The inequality in (\ref{zvar}) is a consequence of  the adiabatic approximation   wherein the physical wavelengths are much smaller than the Hubble radius ($\propto$ the particle horizon). Next, we write $\eta'' = \eta\,\Big[ 1- (\eta-\eta'')/\eta\Big] $ and introduce
\be \omega^{(1)}_k(\eta)\,(\eta-\eta'') = x ~~;~~  \omega^{(1)}_k(\eta)\,(\eta-\eta') = \tau \,, \label{ttau}\ee in terms of which
\be a(\eta'') = a(\eta) \Big[1-\frac{x}{z} \Big] ~~;~~ a(\eta') = a(\eta) \Big[1-\frac{\tau}{z} \Big] \,.\label{aas}\ee This relation allows us to write
\be \big( \omega^{(1)}_k(\eta'') \big)^2 = \big( \omega^{(1)}_k(\eta) \big)^2 + m^2_1 a^2(\eta) \Big[ \Big(1-\frac{x}{z} \Big)^2-1  \Big]  = \big( \omega^{(1)}_k(\eta) \big)^2 \,R^2[x] \,, \label{simpfreq}\ee
 where we introduced
\be R[x;\eta]= \Bigg[1- \frac{2\,x}{\gamma^2_k(\eta) \,z}\,\Big(1-\frac{x}{2z} \Big) \Bigg]^{1/2}\,,   \label{Rvar} \ee with the local Lorentz   factor   given by
\be \frac{1}{\gamma_k(\eta)} = \frac{m_1\,a(\eta)}{ \omega^{(1)}_k(\eta)} = \frac{m_1}{E^{(1)}_k(t)} \,. \label{gammav2}\ee  During (RD) the Lorentz factor can be written as
\be \gamma_k(\eta) = \Bigg[\Big(\frac{a_{nr}}{a(\eta)}\Big)^2+1 \Bigg]^{1/2} = \Bigg[\Big(\frac{\eta_{nr}}{\eta}\Big)^2+1 \Bigg]^{1/2} ~~;~~ \eta_{nr} = \frac{k}{m_1\,H_R}\equiv \frac{a_{nr}}{H_R}\,, \label{gammaRD}\ee the conformal time $\eta_{nr}$ determines the time scale at which the parent particle transitions from relativistic $\eta \ll \eta_{nr}$ to non-relativistic $\eta \gg \eta_{nr}$. In the following analysis we suppress the $\eta$-dependence of $\gamma_k,z$ for simplicity.

We emphasize that the relations (\ref{aas},\ref{simpfreq}) are \emph{exact} in a radiation dominated cosmology. Changing integration variables from $\eta''$ to $x$ given by (\ref{ttau}) and using the above variables we find that the integral (\ref{intIk}) simplifies to the following expression
\be J_k[\eta,\eta'] \equiv J_k[\tau;\eta] = \int^{\tau}_0  \Bigg[1- \frac{2\,x}{\gamma^2_k\,z}\,\Big(1-\frac{x}{2z} \Big) \Bigg]^{1/2}   dx \,,\label{newIk} \ee  obtaining

\be   J_k[\tau;\eta]     =    \tau + \delta_k(\tau;\eta)\,, \label{Isplit}\ee where $\delta_k(\tau)$ is of adiabatic order $\geq 1$ and given by
\be  \delta_k(\tau;\eta)  =   \frac{z}{2} \Bigg\{\Big(1-\frac{2\,\tau}{z} \Big)- \Big(1-\frac{\tau}{z} \Big)\, R[\tau;\eta]  \Bigg\}      \\
 - \frac{z}{2\gamma_k}\,(\gamma^2_k-1) \,\ln \Bigg[\frac{\gamma_k\, R[\tau;\eta] + \big(1-\frac{\tau}{z} \big)}{1+\gamma_k }  \Bigg] \,, \label{deltada}\ee where we recall that both $z$ and $\gamma_k$ depend explicitly on $\eta$.
  Inserting these results into (\ref{Sfunc},\ref{Pfac},\ref{Afac}), and using the new variables $z,\tau$ given by eqns. (\ref{zvar},\ref{ttau}) we find
 \be \mathcal{S}(\eta) = \int^{z}_0 P[\tau;\eta]\, \frac{\sin[A(\tau;\eta)]}{\tau} \, d\tau \,, \label{Sfin}\ee where
 \be P[\tau;\eta] = \frac{\Big[1-\frac{\tau}{z} \Big]}{\sqrt{R[\tau;\eta]}} \,,\label{poftau} \ee and
 \be A[\tau;\eta] = \tau\,\Bigg[1- \Big(1-\frac{1}{\gamma^2_k}\Big)^{1/2}\,  \Bigg] + \delta_k(\tau;\eta)\,, \label{aoftau}\ee where the term in the bracket follows from $k/\omega^{(1)}_k(\eta) = (1-1/\gamma^2_k)^{1/2}$. The  expression (\ref{Sfin}) is amenable to straightforward numerical analysis. However, before we pursue such study, it is  important to recognize several features that will yield to a simplification in the general case of massive daughters. The various factors above display a hierarchy of (dimensionless) time scales  widely separated by $1/z \ll 1$ in the adiabatic approximation: the ``fast'' scale $\tau$, the ``slow'' scale $ \tau/z$ etc.  It is straightforward to find that
 \be \delta_k(\tau;\eta) = - \frac{\tau^2}{2\gamma^2_k\,z}\,   + \cdots \,, \label{deltafiord}\ee confirming that $\delta_k$ is of first and higher adiabatic order. This has a simple, yet illuminating interpretation in terms of an adiabatic expansion of the integral (\ref{intIk}).
 If the frequencies $\omega^{(1)}_k$ were independent of time, this integral would simply be $J_k(\eta,\eta') = \omega^{(1)}_k\,(\eta-\eta') \equiv \tau$. Therefore an expansion of $J_k[\eta,\eta']$ around $\eta'=\eta$ would necessarily entail derivatives of $\omega^{(1)}_k$, namely terms of higher adiabatic order. Consider such an expansion:
   \bea J_k[\eta,\eta'] & = &   0 + \frac{d}{d\eta'}\,J_k[\eta,\eta']\Bigg|_{\eta'=\eta} ~(\eta-\eta') + \frac{1}{2} \,\frac{d^2}{d\eta^{'\,2}}\,J_k[\eta,\eta']\Bigg|_{\eta'=\eta}~(\eta-\eta')^2+ \cdots \nonumber \\
   & = &    \omega^{(1)}_k(\eta)\,(\eta-\eta')- \frac{1}{2} \omega^{'\,(1)}_k(\eta)\, (\eta-\eta')^2 +\cdots \label{adex} \eea
  In terms of $\tau= \omega^{(1)}_k(\eta)\,(\eta-\eta')$, this expansion becomes
 \be J_k[\eta,\eta'] = \tau - \frac{\tau^2}{2\gamma^2_k\,z} +\cdots \label{adex2}\ee  where we used (\ref{adratio}) and (\ref{zvar}). The second term is precisely the leading contribution to $\delta_k$ (\ref{deltafiord}). This analysis makes explicit that an expansion of the integral (\ref{intIk}) in powers of $\eta-\eta'$ is precisely an adiabatic expansion in terms of derivatives of the frequencies. Since the n-th power of $\eta-\eta'$ in such expansion is multiplied by the $n-1$ derivative of the frequencies, and when $(\eta-\eta')$ is replaced by $\tau/\omega^{(1)}_k(\eta)$ the $n-1$ derivative of the frequencies is divided by $(\omega^{(1)}_k(\eta))^n$ yielding a dimensionless ratio of adiabatic order $n-1$.

 Let us now consider the full integral expression for $\mathcal{S}(\eta)$ given by (\ref{Sfin}) with the corresponding expressions for $P[\tau]$ and $\delta_k(\tau)$. For $z \gg 1$ the terms of the form $\tau/z, \tau^2/z^2$ will be negligible in most of the integration region but for the region of $\tau \approx z$ where these terms become of $\mathcal{O}(1)$. However, because of the factor $\tau$ in the denominator of the integrand in (\ref{Sfin}), a consequence of the momentum integration, this region is suppressed by a factor $1/z \ll 1$ yielding effectively a contribution of first (and higher) adiabatic order. Therefore the contribution from adiabatic corrections, proportional to powers of $\tau/z$ are, in fact, subleading. This argument suggests that the \emph{zeroth order} adiabatic approximation to $\mathcal{S}(\eta)$, namely
 \be \mathcal{S}_0(\eta) = \frac{2}{\pi}\,\int^z_0  \, \frac{\sin[A_0(\tau;\eta)]}{\tau} \, d\tau ~~;~~A_0[\tau;\eta] = \tau\,\Bigg[1- \Big(1-\frac{1}{\gamma^2_k}\Big)^{1/2}\,  \Bigg]  \,, \label{Sad}\ee should be a very good approximation to the full function  $\mathcal{S}(\eta)$ for $z\gg 1$ with closed form expression
 \be  \mathcal{S}_0(\eta) = \frac{2}{\pi}\,Si[A_0(z(\eta);\eta)]\,.  \label{S0Si} \ee where $Si[x]$ is the sine-integral function with asymptotic behavior $Si[x] \rightarrow \pi/2 - \cos(x)/x +\cdots$ as $x\rightarrow \infty$. This function rises and begins to oscillate around its asymptotic value at $x \simeq \pi$. This behavior implies that the rise-time of $Si[A_0(z;\eta)]$ to its asymptotic value in the ultrarelativistic case $\gamma_k \gg 1$ increases $\propto \gamma^2_k$. In fact one finds that the full function $\mathcal{S}(\eta)$ and its first order adiabatic approximation $\mathcal{S}_0(\eta)$ \emph{vanish} as $\gamma_k \rightarrow \infty$. Namely, the contribution from $\mathcal{S}_0$ (and similarly from $\mathcal{S}$) is negligible while the particle is ultrarelativistic.  This expectation is verified numerically.

\begin{figure}[ht!]
\begin{center}
\includegraphics[height=4in,width=3.2in,keepaspectratio=true]{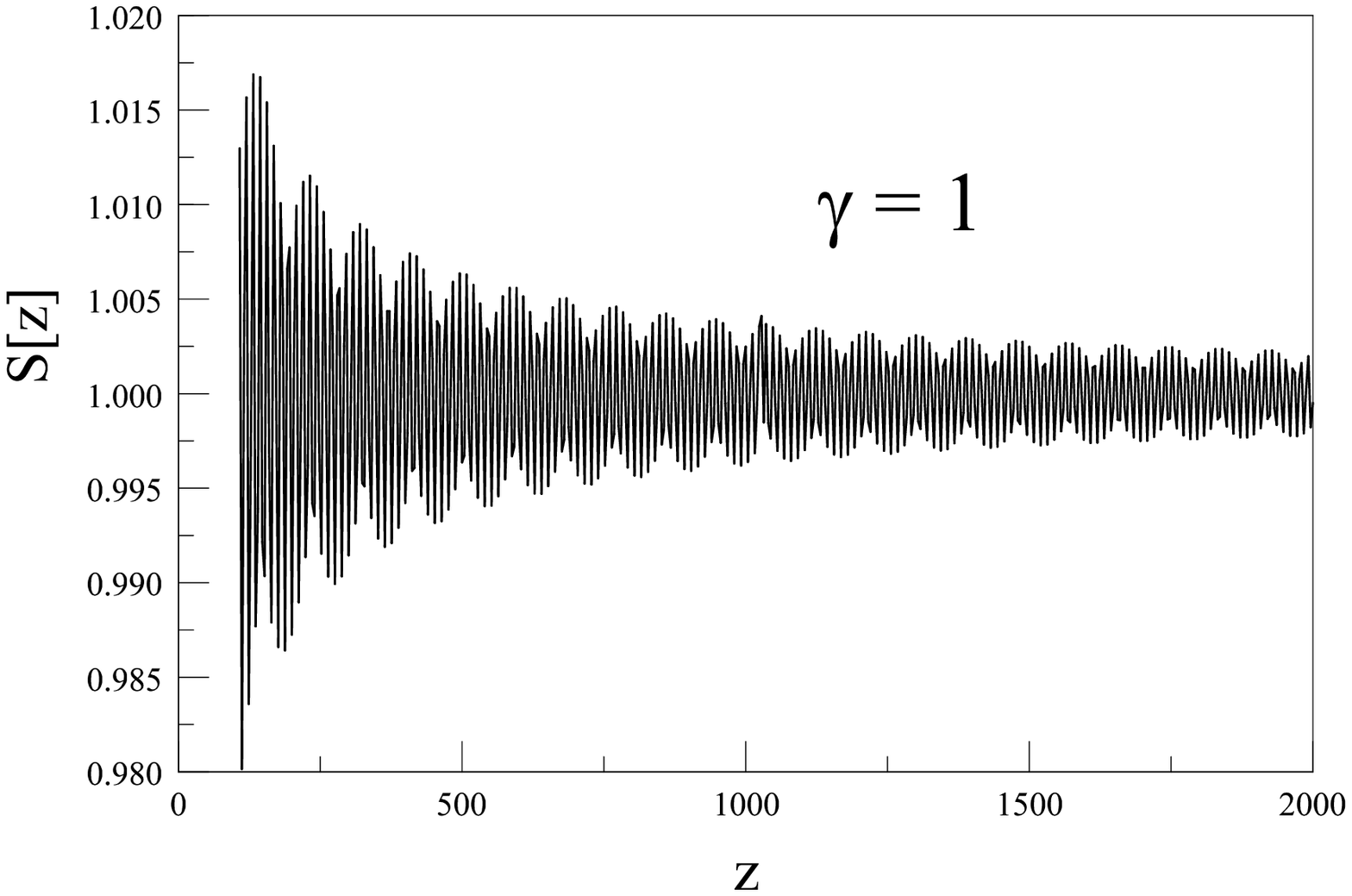}
\includegraphics[height=4in,width=3.2in,keepaspectratio=true]{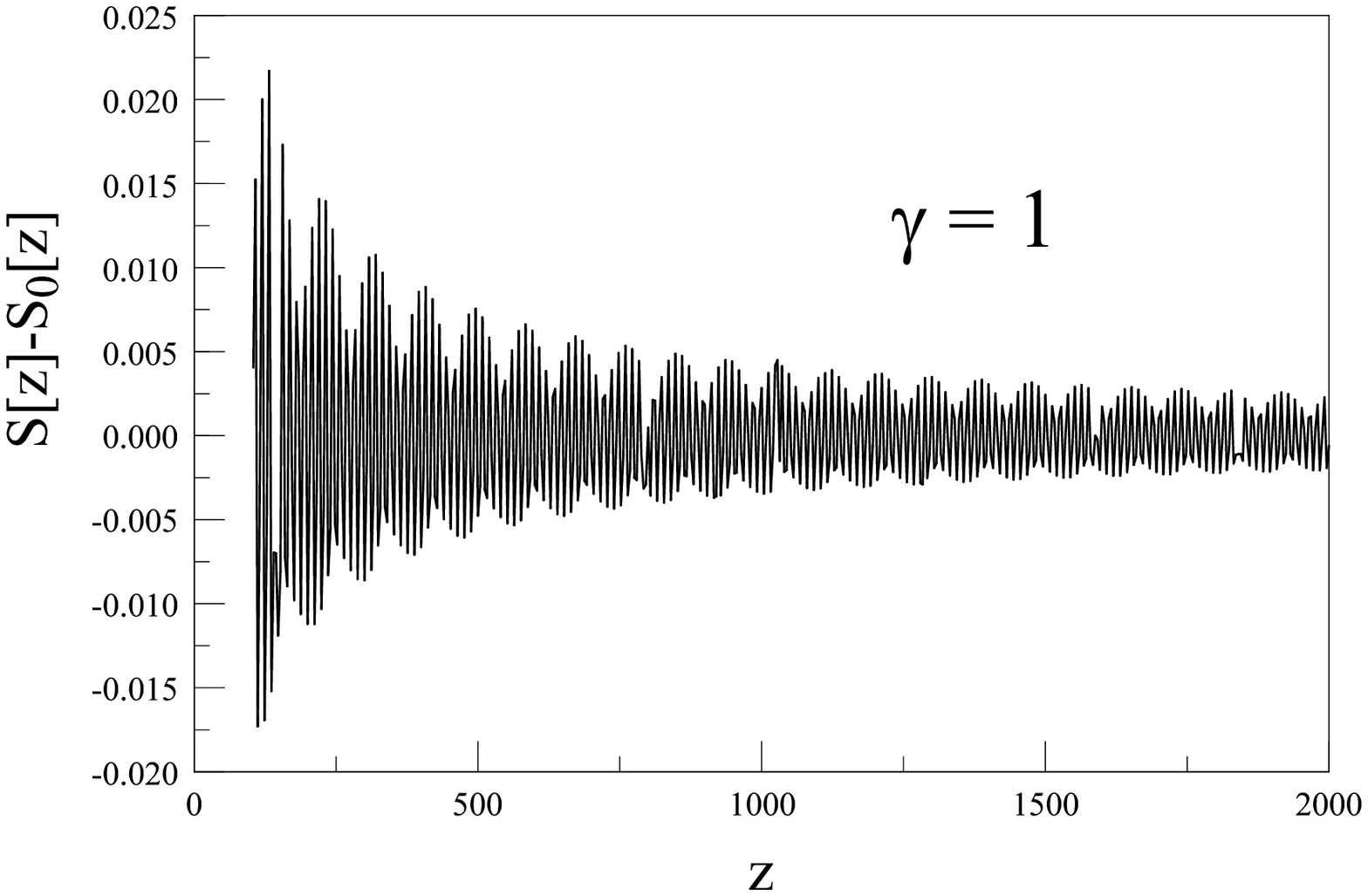}
\caption{$S[z]$ and $S[z]-S_0[z]$ vs. $z$ for $\gamma_k =1$. }
\label{fig4:sgamma1}
\end{center}
\end{figure}

\begin{figure}[ht!]
\begin{center}
\includegraphics[height=3.5in,width=3.2in,keepaspectratio=true]{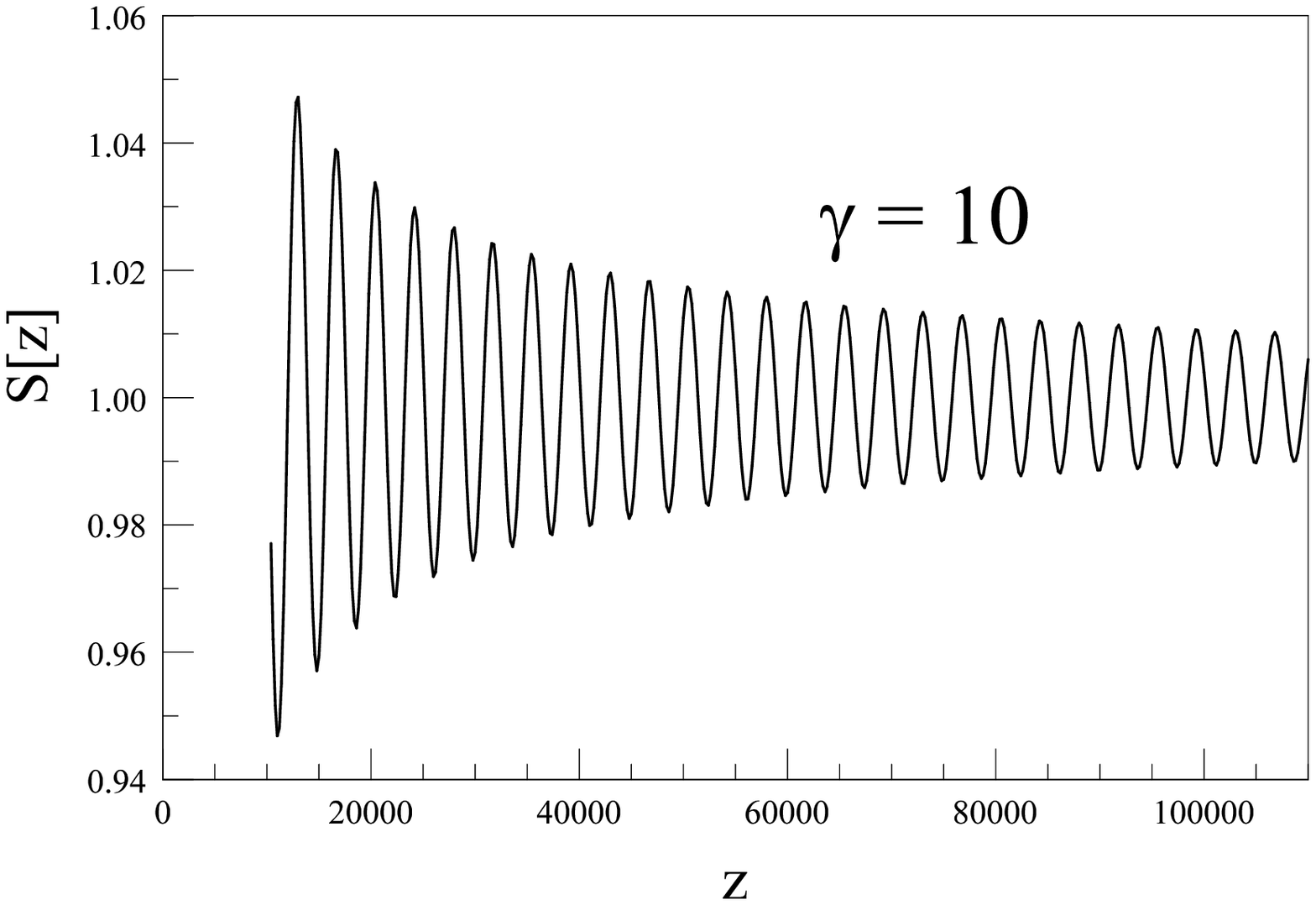}
\includegraphics[height=3.5in,width=3.2in,keepaspectratio=true]{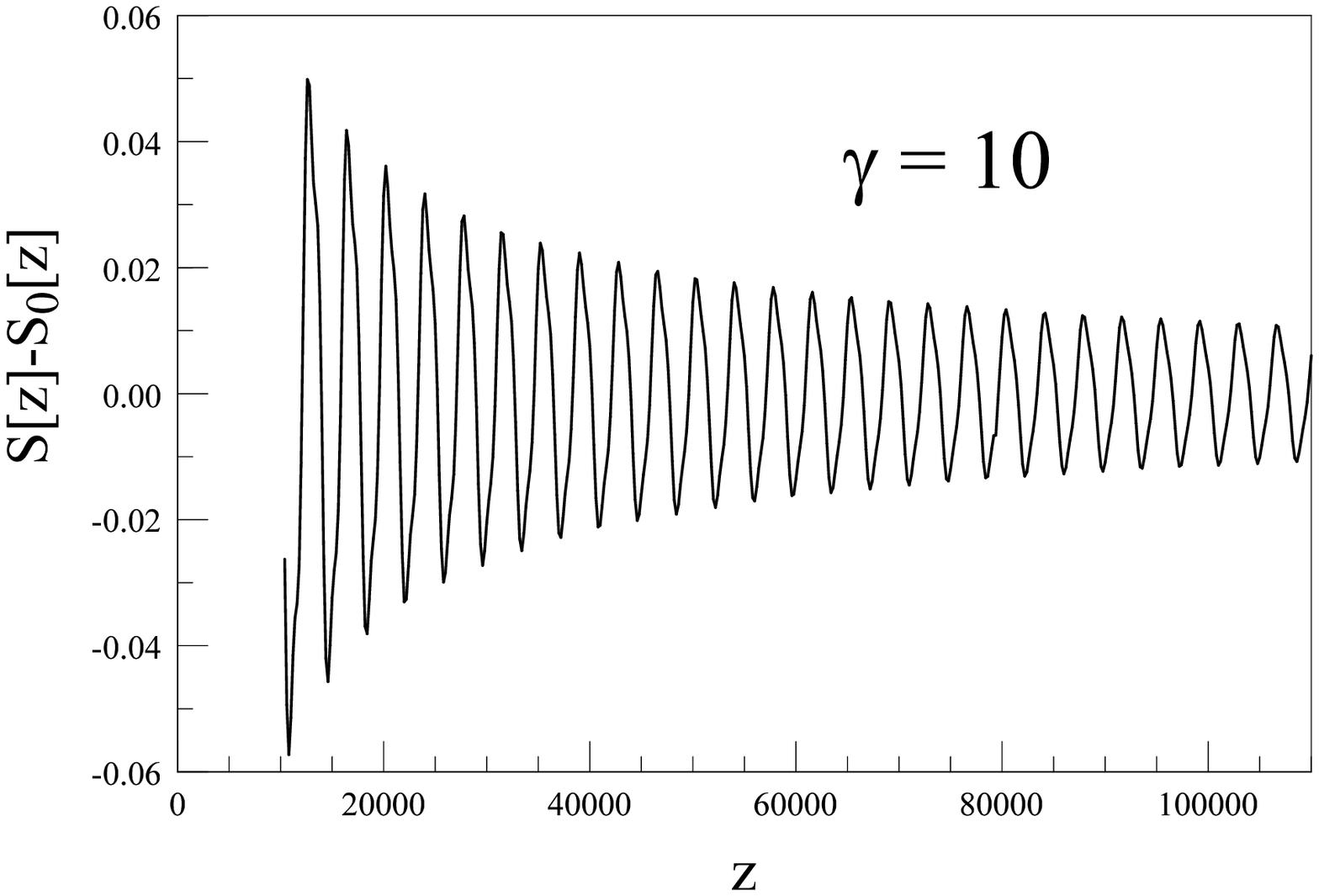}
\caption{$S(z)$ and $S(z)-S_0(z)$ vs. $z$ for $\gamma_k =10$. }
\label{fig5:sgamma10}
\end{center}
\end{figure}

Figures (\ref{fig4:sgamma1},\ref{fig5:sgamma10}) display $S(z)$ and $S(z)-S_0(z)$ vs. $z$ for the non-relativistic limit $\gamma_k=1$ and for the relativistic regime $\gamma_k=10$. In both cases these figures confirm that the zeroth adiabatic approximation $S_0(z)$ is   excellent   for $z \gg 1$. They also confirm the slow rise of this contribution in the ultrarelativistic case, note the scale on the horizontal axis for the case $\gamma_k =10$ compared to that for $\gamma_k=1$. For $\gamma_k > 1$ we have displayed the results for a fixed value of $\gamma_k$ to illustrate the main behavior for the non-relativistic and relativistic limits and highlight that the relativistic case features a larger rise-time. Obviously a detailed numerical study including the $\eta$ dependence of $\gamma_k$ will depend on the particular values of $k,m_1$.

Replacing the function $\mathcal{S}(\eta)$ by the zeroth order approximation $S_0(\eta)$ is also consistent with our main approximation of keeping only the zeroth order adiabatic contribution in the mode functions. Therefore consistently with the zeroth adiabatic order, we find that the decay rate for the case of a massive parent decaying into two massless daughters is given by
\be   \Gamma_k(\eta) = \frac{\lambda^2\,a^2(\eta)}{8\pi \,\omega^{(1)}_k(\eta)}\,\,\frac{1}{2}\Bigg[1+ \frac{2}{\pi}\,Si[A_0(z(\eta);\eta)]\Bigg]~~;~~A_0(z(\eta);\eta) = z(\eta)\,\Bigg[1- \Big(1-\frac{1}{\gamma^2_k(\eta)}\Big)^{1/2}\,  \Bigg] \,. \label{Gamav2}\ee

 We emphasize that although in several derivations leading up to the results (\ref{Sfin},\ref{poftau},\ref{aoftau}) we have used the scale factor during the RD dominated era, for example in eqns. (\ref{aas},\ref{simpfreq}), only the explicit dependence of $\delta_k(\tau,\eta)$  and the prefactor $P[\tau;\eta]$ on $\tau,\eta$ depend  on this choice. However, as shown above the leading adiabatic order corresponds to taking $\delta_k =0 $ and $P[\tau,\eta]=1$, namely $\delta_k$ and  the $\tau,\eta$ dependent terms in $P[\tau,\eta]$ yield contributions of higher adiabatic order. Therefore,   the leading (zeroth) adiabatic order given by (\ref{Gamav2})  is valid either for the (RD) or (MD) dominated eras.

  Remarkably, this result is similar to that in Minkowski space time obtained in appendix (\ref{app:Mink}) with the \emph{only} difference being the   scale factor    and explicit time dependence of the frequency.

The \emph{decay law} of the probability, given by (\ref{probaA}) requires the integral of the rate (\ref{Gamav2}). It is now convenient to pass to comoving time in terms of which we find (again setting $\eta_i=0$)
\be \int^\eta_0 \Gamma_k(\eta) ~ d\eta \equiv \Gamma_0  \int^t_0 \frac{\mathcal{F}(t')}{\gamma_k(t')} ~ dt' \,, \label{decaylaw}\ee where
\be \Gamma_0 = \frac{\lambda^2}{8\pi m_1}~~;~~ \mathcal{F}(t')= \frac{1}{2}\Big[1+ \frac{2}{\pi}\,Si[A_0(t')]\Big] \,,\label{decaylawdefs}\ee  where $\Gamma_0 $ is the decay rate of a particle at rest in Minkowski space-time and $\gamma_k(t)$ the time dilation factor, which depends explicitly on time   as a consequence of the cosmological redshift of the physical momentum.

 Up to  the factor $\mathcal{F}(t')$, the decay rate in comoving time has a simple interpretation:
 \be \Gamma_k(t) \simeq \frac{\Gamma_0}{\gamma_k(t)}\,,  \label{ratecos}\ee namely a decay width at rest divided by the time dilation factor. During (RD) it follows that
 \be \gamma_k(t) = \Big[1+ \frac{t_{nr}}{t} \Big]^{1/2}~~;~~ t_{nr} = \frac{k^2}{2m^2_1H_R} \,, \label{gamakoftRD}\ee  where $t_{nr}(k)$ is the transition time scale between the ultrarelativistic ($t\ll t_{nr}$)  and non-relativistic ($t\gg t_{nr}$) regimes, assuming that the transition occurs during the (RD) era, which is a suitable assumption for masses larger than a few $\mathrm{eV}$.

In the (RD) era we find (using \ref{zvar}, \ref{gammav2}, \ref{gammaRD}, and \ref{Gamav2})
\bea z(t) & = & \Bigg[\frac{k^2}{m_1\,H_R}  \Bigg] \, \Bigg[\frac{t}{t_{nr}}\Big(1+\frac{t}{t_{nr}} \Big) \Bigg]^{1/2}\,, \label{zoftt}\\
 A_0(t)  & = &   \Bigg[\frac{k^2}{m_1\,H_R}  \Bigg] \, \sqrt{\frac{t}{t_{nr}}}\,\Bigg[  \Big(1+\frac{t}{t_{nr}}  \Big)^{1/2}-1  \Bigg]  \,. \label{A0oft}\eea

  In Minkowski space time, the calculation of the decay rate in S-matrix theory takes the initial and final states at $t = \mp \infty$ respectively, in which case the $Si$ function attains its asymptotic value and $\mathcal{F}=1$. The derivation of the decay rate in Minkowski space-time but in real time implementing the Wigner-Weisskopf method is described in detail in appendix (\ref{app:Mink}) and offers a direct comparison between the flat and curved space time results.

 In general the integral in (\ref{decaylaw})  must be obtained numerically. However, in order to understand the main differences resulting from the cosmological expansion we first focus on the non-relativistic  and the ultra-relativistic limits respectively.
\vspace{2mm}

\textbf{Non-relativistic limit:}

\vspace{2mm}

In this limit we set $k=0$ with $\gamma_k(t) = 1$ and for simplicity we take the $Si$ function to have reached its asymptotic value, therefore replacing $\mathcal{F}(t') \simeq 1$ inside the integrand yielding\footnote{Keeping the function $\mathcal{F}$ in the integrand yields a subdominant constant term in the long time limit. A similar term is found in ref.\cite{vilja}.}

\be \int^\eta_0 \Gamma_{k=0}(\eta') ~ d\eta' = \frac{\lambda^2 }{8\pi m_1}\, t \,. \label{decaylawNR}\ee
This is precisely the decay law in Minkowski space time and coincides with the results obtained in ref.\cite{vilja}. However this is the case \emph{only} if the parent particle is ``born'' at rest in the comoving frame, otherwise the time dilation factor modifies (substantially, see below) the decay rate and law.

\vspace{2mm}

\textbf{Ultra-relativistic limit:}

\vspace{2mm}

In this limit we set $m_1=0$ corresponding to $\gamma_k \rightarrow \infty$ in the argument of the $Si$ function, in which case its contribution vanishes identically,  yielding $\mathcal{F}(t') = 1/2$ and

 \be \int^\eta_0 \Gamma_k(\eta) ~ d\eta \equiv \frac{\lambda^2}{16\pi} \int^t_0 \frac{1}{k_p(t')} ~ dt' \,, \label{URdelaw} \ee  with  \emph{physical} wavevector $k_p(t) = k/a(\eta(t))$. During (RD) this result  yields the following decay law of the probability
\be \Big|C^{(1)}_{\vk}(t)\Big|^2  =  e^{-(t/t^*)^{3/2}} ~~;~~ t^* = \Bigg[\frac{\lambda^2~(2 H_R )^{1/2}}{24\pi\, k}\Bigg]^{-2/3}\, \,.  \label{probUR}\ee This decay law is a \emph{stretched exponential}, it is a distinct  consequence of time dilation combined with  the cosmological redshift of the physical momentum.

Although obtaining the decay law throughout the full range of time entails an intense numerical effort and depends in detail on the various parameters $k,m_1,H_R$ etc. We can obtain an approximate but more clear understanding of the transition between the ultrarelativistic and non-relativistic regimes by focusing solely on  the time integral of the inverse Lorentz factor, because the contribution from $\mathcal{F}$ is bound $1/2 \leq \mathcal{F} \leq 1$. Therefore, setting $\mathcal{F}=1$ and during (RD) we find
\bea \int^t_0 \Gamma_k(t')\,  dt' &  =  & \Gamma_0 \,t_{nr}\, {G}_k(t)  \nonumber \\
 {G}_k(t) & = & \Bigg[\frac{t}{t_{nr}}\Big(1+\frac{t}{t_{nr}} \Big)  \Bigg]^{1/2} - \ln\Bigg[\sqrt{1+\frac{t}{t_{nr}}}+\sqrt{\frac{t}{t_{nr}}} \Bigg]  \,.  \label{gammainter}\eea
    For the ultrarelativistic regime $t\ll t_{nr}$ we find the result (\ref{probUR}) up to a factor $1/2$ because we have set $\mathcal{F}=1$, whereas in the non-relativistic regime,  for $t \gg t_{nr}$, we obtain the transition probability
  \be  \Big|C^{(1)}_{\vk}(t)\Big|^2   = e^{-\Gamma_0\,t}\, \Big(\frac{t}{t_{nr}}\Big)^{\Gamma_0 t_{nr}/2} \,, \label{powlaw}  \ee again,  the extra power of time is a consequence of the cosmological redshift in the time dilation factor. For $k=0$, namely $t_{nr}=0$,  we obtain the non-relativistic result (\ref{decaylawNR}).

  The function $ {G}_k(t)$ interpolates between the ultrarelativistic case $\propto t^{3/2}$ for $t \ll t_{nr}$ and the non-relativistic case $\propto t$ for $t \gg t_{nr}$ and encodes the time dependence of the time dilation factor through the cosmological redshift.

\begin{figure}[ht!]
\begin{center}
\includegraphics[height=4in,width=3.2in,keepaspectratio=true]{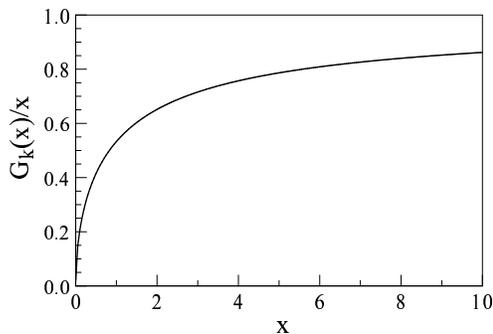}
\caption{The ratio $G_k(x)/x$ for $x=t/t_{nr}$.}
\label{fig:ratiodecay}
\end{center}
\end{figure}

 In Minkowski space time  the   result of the integral in (\ref{gammainter}) is simply $\Gamma_0 t$ which is conveniently written as  as $\Gamma_0 t_{nr}\, (t/t_{nr})$. Because $G_k$ is a function of $t/t_{nr}$,  a measure of the \emph{delay} in the cosmological decay compared to that of Minkowski space time is given by the ratio $G_k(x)/x$ with $x\equiv t/t_{nr}$. This ratio is displayed in fig. (\ref{fig:ratiodecay}), it vanishes as $x\rightarrow 0$ as $x^{1/2}$ and $G_k(x)/x  \rightarrow 1$  as $x\rightarrow \infty$, in particular $G_k(1)=\sqrt{2}-\ln[1+\sqrt{2}]=0.533$.

This analysis suggests that the effect of the cosmological expansion can be formally included by defining a  \emph{time dependent effective decay rate},
\be \widetilde{\Gamma}_k(t) = \Gamma_0\,(G_k(x)/x)~~;~~x=t/t_{nr}\,, \label{effgama}\ee  and $t_{nr}$ depends on k (see (\ref{gammainter})), so that the decay law for the survival probability of the parent particle  becomes
\be  P(t)= e^{-\widetilde{\Gamma}_k(t)\,t}\,. \label{surviprob}\ee This effective decay rate is \emph{always} smaller than the Minkowski rate for $k\neq 0$ as a consequence of time dilation and its time dependence through the cosmological redshift, coinciding with the Minkowski rate at rest only for $k=0$, namely particles born and decaying at rest in the comoving frame.

Finally, the effect of the function $\mathcal{F}$ must be studied numerically for a given set of parameters $k,m_1$. However, we can obtain an estimate during the (RD) era from the expression (\ref{A0oft}) for the argument of the $Si$-function. Writing
\be \Bigg[\frac{k^2}{m_1\,H_R}\Bigg] \equiv  \beta \simeq  10^{16}\,\Bigg[ \frac{\Big(k/10^{-13}\,\mathrm{GeV}\Big)^2}{\Big(m_1/100\,\mathrm{GeV}\Big)}\Bigg] \,,\label{prefa}\ee it follows that $A_0(t) \ll 1$ for $t/t_{nr} \ll 1/ \beta^{2/3}$ and $A_0(t) > 1$ for $t/t_{nr} > 1/ \beta^{2/3}$. Because $Si[x] \propto x$ as $x \rightarrow 0$ and approaches $\pi/2$ for $x \simeq \pi$  the large pre-factor in (\ref{prefa}) for typical values of $k,m_1$ entails that   the transition between these regimes is fairly sharp, therefore we can approximate the function $\mathcal{F}(t')$ as
\be \mathcal{F}(t') \approx \frac{1}{2}\,\Theta\Big(\beta^{-2/3}-t'/t_{nr}\Big)+\Theta\Big(t'/t_{nr}-\beta^{-2/3} \Big)\,.\label{Fappx} \ee

\subsection{Massive parent and daughters}\label{sub:allmass}

 We now consider the self energy (\ref{sigadzero}) for the case of massive daughters. Unlike the case of massless daughters, in this case neither the time nor the momentum integrals can  be done analytically. However, the study of  massless daughters revealed that  the adiabatic approximation  in   the time integrals is excellent when the adiabatic conditions $H(t)/E_k(t)\ll 1$ are fulfilled for all species. The analysis of the previous section has shown that inside the time integrals we can replace $a(\eta') \rightarrow a(\eta)~;~\omega_k(\eta') \rightarrow \omega_k(\eta)$ since the difference is at least first order (and higher) in the adiabatic approximation (see the results for the  factor $P(\tau)$ in eqn. (\ref{Sfin})). Furthermore, carrying an adiabatic expansion of the time integrals of the frequencies is tantamount to expanding these in powers of $\eta-\eta'$, with the first term, proportional to $\eta-\eta'$ yielding the zeroth adiabatic order and the higher powers of $\eta-\eta'$ being of higher adiabatic order. Replacing $\eta-\eta'= \tau/\omega^{(1)}_k(\eta)$ associates the higher powers of $\tau$ with higher orders in the adiabatic expansion as discussed above. However, this argument by itself does not guarantee the reliability of the adiabatic expansion because for $\tau \simeq z = E_k/H$ each term in this expansion becomes of the same order. What guarantees the reliability of the adiabatic expansion is the momentum integral that suppresses the large $\eta-\eta'$ regions. This is manifest in the $1/\tau$ suppression of the integrand in the case of massless daughters (see eqn. (\ref{Sfin})). This can be understood from a simple observation. Consider the momentum integral in the full expression (\ref{sigadzero}), setting $\eta=\eta'$ in the exponent yields a linearly divergent momentum integral. This is the origin of the singularity as $\eta \rightarrow \eta'$ in (\ref{integI}).  The contributions from regions with large $\eta-\eta'$ oscillate very fast and are suppressed. Therefore the momentum integral is dominated by the region of small $\eta-\eta'$. In  appendix (\ref{app:validity}) we provide an analysis of  the first adiabatic correction and confirm both analytically and numerically that it is indeed suppressed by the momentum integration also in the case of massive daughters.

 Therefore we consider the  leading adiabatic order that yields
 \be \Gamma_k(\eta) =     \frac{2\,\lambda^2\,a^2(\eta)}{ \omega^{(1)}_k(\eta)} \int \frac{d^3p}{(2\pi)^3} \frac{1}{2\,\omega^{(2)}_p(\eta)\,2\,\omega^{(2)}_q(\eta)}~
 \frac{\sin\Big[\Big(\omega^{(1)}_k(\eta)-\omega^{(2)}_p(\eta)-\omega^{(2)}_q(\eta) \Big)\,\eta  \Big]}{\Big(\omega^{(1)}_k(\eta)-\omega^{(2)}_p(\eta)-\omega^{(2)}_q(\eta) \Big)} ~~;~~ q = |\vk-\vp|\,.  \label{massidau} \ee It is convenient to recast this expression in terms of the physical (comoving) energy and momenta: $\omega_k(\eta)= a(\eta) E_k(t)$   absorbing the three powers of $a(\eta)$  in the denominator in the momentum integral in (\ref{massidau}) into the measure $d^3 p \rightarrow d^3 p_{ph}$ where $p_{ph}(\eta) \equiv p/a(\eta)$ is the physical momentum, keeping the same notation for the integration variables (dropping the subscript ``ph'' from the momenta)  to simplify notation, we obtain
 \be \Gamma_k(\eta) =     \frac{2\,\lambda^2\,a(\eta)}{ E^{(1)}_k(\eta)} \int \frac{d^3p}{(2\pi)^3} \frac{1}{2\,E^{(2)}_p(\eta)\,2\,E^{(2)}_q(\eta)}~
 \frac{\sin\Big[\Big(E^{(1)}_k(\eta)-E^{(2)}_p(\eta)-E^{(2)}_q(\eta) \Big)\,\widetilde{T} \Big]}{\Big(E^{(1)}_k(\eta)-E^{(2)}_p(\eta)-E^{(2)}_q(\eta) \Big)} ~~;~~ q = |\vk-\vp|   \,.  \label{massidauph} \ee The variable
 \be \widetilde{T}   =  a(\eta)\,\eta \equiv \frac{1}{\widetilde{H}} =    \Bigg\{  \begin{array}{c}
                                \frac{1}{H}~~(RD) \\
                                 \frac{2}{H}~~(MD)
                              \end{array} \,,\label{bigT}\ee
corresponds to the physical particle horizon, proportional to the Hubble time. Obviously the momentum integrals  cannot be done in closed form, however (\ref{massidauph}) becomes more familiar with a dispersive representation, namely
\be \Gamma_k(\eta) = \int^\infty_{-\infty}  dk_0 ~  \rho(k_0,k) \, \frac{\sin\Big[\big(k_0 - E^{(1)}_k(\eta) \big) \widetilde{T} \Big]}{\pi \big(k_0 - E^{(1)}_k(\eta) \big) } \,, \label{disp} \ee with the spectral density
\be \rho(k_0,k;\eta) =  \frac{\lambda^2\,a(\eta)}{ E^{(1)}_k(\eta)} \int \frac{d^3p}{(2\pi)^3} \frac{(2\pi)\, \delta\Big[k_0-E^{(2)}_p(\eta)-E^{(2)}_q(\eta)  \Big] }{2\,E^{(2)}_p(\eta)\,2\,E^{(2)}_q(\eta)} \,, \label{rhok0k} \ee we refer to (\ref{disp}) the \emph{cosmological Fermi's Golden Rule}. In the formal limit $\widetilde{T} \rightarrow \infty$
\be  \frac{\sin\Big[\big(k_0 - E^{(1)}_k(\eta) \big) \widetilde{T} \Big]}{\pi \big(k_0 - E^{(1)}_k(\eta) \big) } ~~ \longrightarrow  ~~ \delta(k_0 - E^{(1)}_k(\eta))\,. \label{longtime} \ee The density of states (\ref{rhok0k}) is the familiar two body decay phase space in Minkowski space-time for a particle of energy $k_0$ into two particles of equal mass. It is given by (see appendix (\ref{app:Mink})),
\be \rho(k_0,k;\eta) =\frac{\lambda^2\,a(\eta) }{8\pi\,E^{(1)}_k(\eta)} \, \Bigg[1- \frac{4\,m^2_2}{k^2_0-k^2} \Bigg]^{1/2} \,\Theta(k^2_0 - k^2-4\,m^2_2)\,\Theta(k_0)  \,, \label{rhok0kcos}\ee where $k \equiv k_{ph}(\eta) $ is the the physical momentum, and the theta function describes  the reaction threshold.

Before we proceed to the study of $\Gamma_k(\eta)$ for $m_2\neq 0$, we establish a direct connection with the results of the previous section for $m_2=0$, where the momentum integration was carried out first. Setting $m_2=0$ in (\ref{rhok0kcos}), inserting it into the dispersive integral (\ref{disp}) and changing variables $(k_0 - E^{(1)}_k(\eta)) \,\widetilde{T} \rightarrow x $ we find
\be \Gamma_k(\eta) = \frac{\lambda^2\,a(\eta) }{8\pi\,E^{(1)}_k(\eta)} \int^\infty_{-\big(E^{(1)}_k(\eta)-k\big)\,\widetilde{T}}  ~ \frac{\sin(x)}{\pi\,x} ~dx  = \frac{\lambda^2\,a(\eta) }{8\pi\,E^{(1)}_k(\eta)} \,\frac{1}{2}\Bigg[1+\frac{2}{\pi}\,Si\Big[ \big(E^{(1)}_k(\eta)-k\big)\,\widetilde{T}\Big] \Bigg]\,, \label{m2zerocase} \ee which is precisely the result (\ref{Gamav2}) displaying the ``prompt'' $(1)$ and ``raising'' $(Si)$ terms inside the bracket.

Restoring $m_2 \neq 0$, and taking formally the infinite time limit (\ref{longtime}), the rate (\ref{disp}) becomes
\be \Gamma(\eta) = \frac{\lambda^2\,a(\eta) }{8\pi\,E^{(1)}_k(\eta)} \, \Bigg[1- \frac{4\,m^2_2}{m^2_1} \Bigg]^{1/2} \,\Theta(m^2_1-4\,m^2_2)\,,\label{gammalongtime} \ee revealing the usual  two particle threshold $m_1 \geq 2 \, m_2$.

\vspace{2mm}

\textbf{Threshold relaxation:}

\vspace{2mm}

\begin{figure}[ht!]
\begin{center}
\includegraphics[height=4in,width=3.2in,keepaspectratio=true]{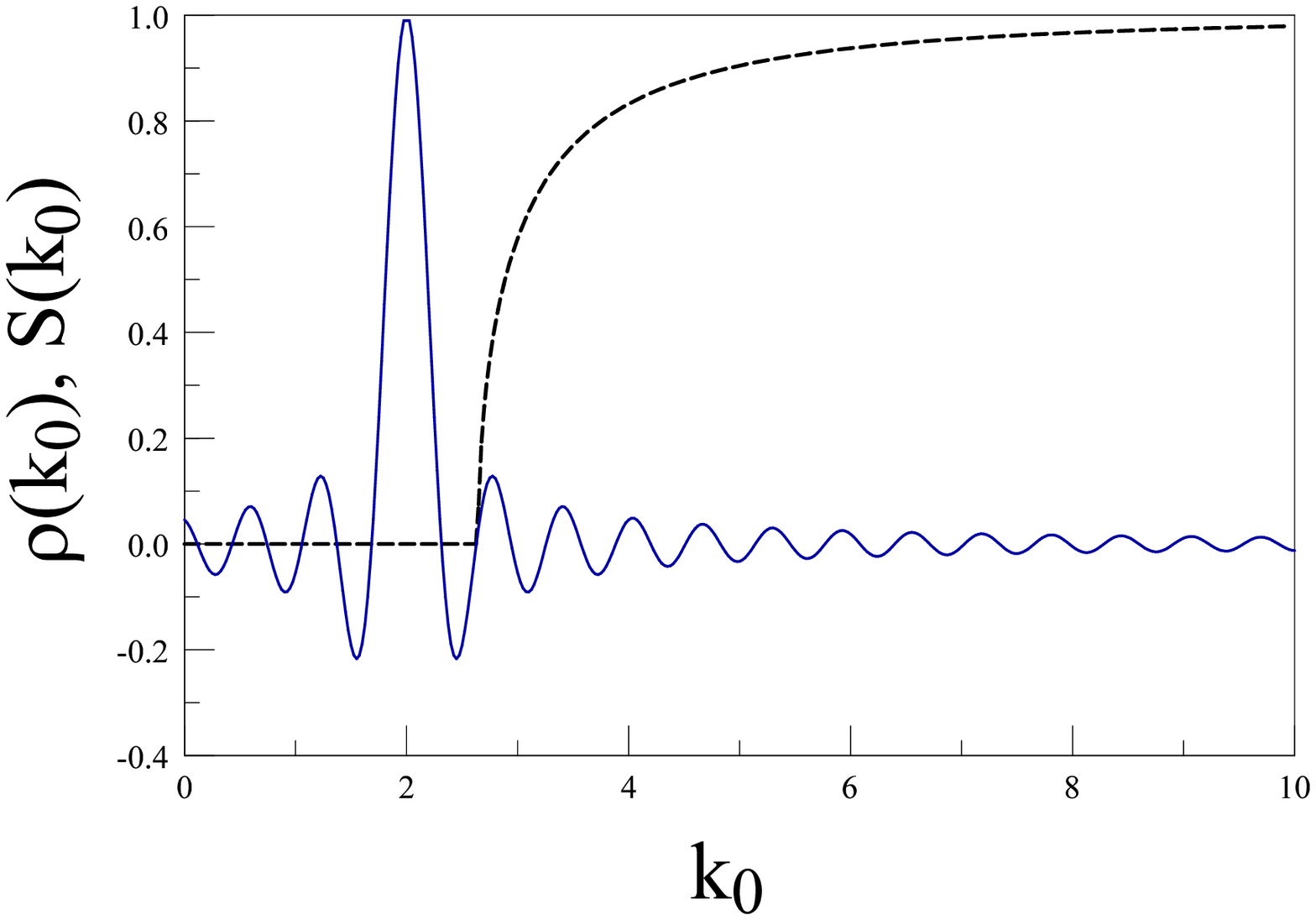}
\includegraphics[height=4in,width=3.2in,keepaspectratio=true]{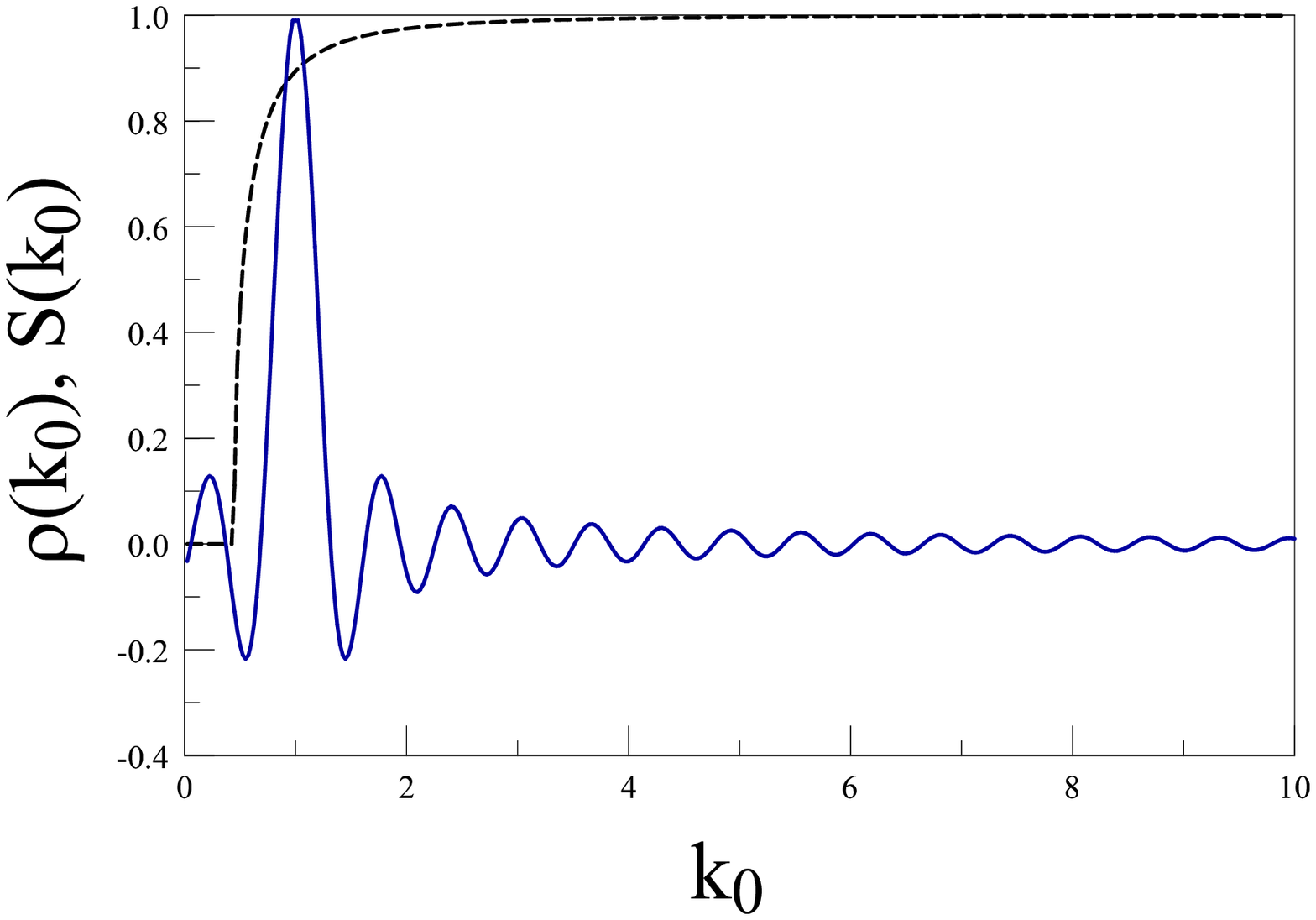}
\caption{The functions $\rho(k_0,k)$ (dashed line) and $S(k_0)= \sin[(k_0-E)T]/[(k_0-E)T]$ in units of $m_1$. Left panel: $E=2, 4m^2_2 =4, T=10$ corresponding to $E$ below threshold. Right panel: $E=1, 4m^2_2 =0.2, T=10$  corresponding to $E$ above threshold.}
\label{fig:thresholds}
\end{center}
\end{figure}

\begin{figure}[ht!]
\begin{center}
\includegraphics[height=4in,width=4in,keepaspectratio=true]{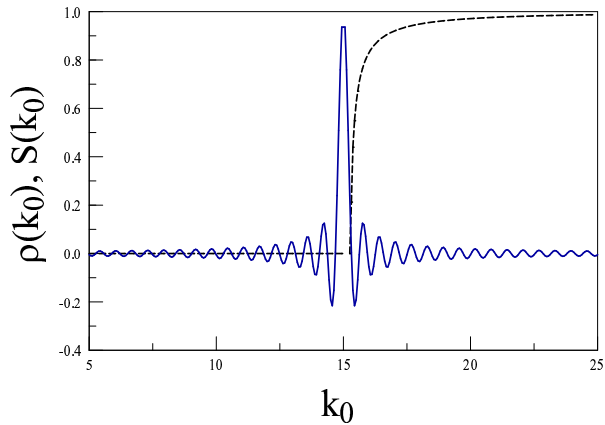}
\caption{The functions $\rho(k_0,k)$ (dashed line) and $S(k_0)= \sin[(k_0-E)T]/[(k_0-E)T]$ in units of $m_1$ for   $E=15, 4m^2_2 =10, T=10$ corresponding to an ultrarelativistic parent with $E$ below threshold.  }
\label{fig:thresholdsbe}
\end{center}
\end{figure}

However, before taking the infinite time limit we recognize important physical consequences of the rate (\ref{disp}). The Hubble time $\widetilde{T}$ introduces an uncertainty in energy $\Delta E \simeq 1/\widetilde{T} \equiv  \widetilde{H}$ which allows physical processes that \emph{violate local energy conservation}
 on the scale of this uncertainty. In particular, this uncertainty allows a particle of mass $m_1$ to decay into \emph{heavier} particles, as a consequence of the relaxation of the threshold condition via the uncertainty. This remarkable feature can be understood as follows. The sine function in  (\ref{disp}) features a maximum at $k_0 = E^{(1)}_k(\eta)$ with half-width (in the variable $k_0$) $\approx \pi \widetilde{H}$, narrowing as $\widetilde{T}$ increases.
 The spectral density $\rho(k_0,k;\eta)$ has support above the threshold at $k^*_0 = \sqrt{k^2+4m^2_2}$, it is convenient to write this threshold as $k^*_0 = \sqrt{(E^{(1)}_k(\eta))^2+ (4m^2_2 -m^2_1)}$.  For $4m^2_2 -m^2_1 <0$ the position of the peak of the sine function, at $k_0 = E^{(1)}_k(\eta)$  lies above the threshold, but for $4m^2_2 -m^2_1 >0$ it lies below it. In this latter case,  if the condition
 \be \Big(E^{(1)}_k(\eta)+\pi \widetilde{H}\Big)^2 \gg (E^{(1)}_k(\eta))^2+ (4m^2_2 -m^2_1) \label{condi}\ee is fulfilled,   the ``wings'' of the sine function beyond the peak feature a large overlap with the region of support of the spectral density. This is displayed in figs. (\ref{fig:thresholds},\ref{fig:thresholdsbe} ). This phenomenon entails the opening of unexpected new  channels  for  a   particle to decay into two \emph{heavier} particles as a consequence of the energy uncertainty determined by the Hubble time.

In the adiabatic approximation with $E^{(1)}_k(\eta) \gg \widetilde{H}$ the overlap condition (\ref{condi}) reads
\be 2\pi \, E^{(1)}_k(\eta) \,\widetilde{H}(\eta) \gg 4m^2_2 -m^2_1\,, \label{condith}\ee which shows that this condition becomes more easily fulfilled for a relativistic parent. This is clearly displayed in fig. (\ref{fig:thresholdsbe}).

To gain better understanding of this condition, let us consider the specific case of an ultrarelativistic parent with mass $m_1 \simeq 100 \,\mathrm{GeV}$ with a GUT-scale comoving energy $E_k \simeq  10^{15}\,\mathrm{GeV}$ decaying into two daughters with mass $m_2 \simeq 1\,\mathrm{TeV}$ for illustration. We can then replace $E_k \simeq k/a(\eta)$ with $k \simeq 10^{-13}\,\mathrm{GeV}$ being the comoving momentum that yields a physical momentum $k_{ph} \simeq  10^{15}\,\mathrm{GeV}$ (with $a(\eta_i) \simeq 10^{-28}$), furthermore with $\widetilde{H} \simeq H_R/a^2(\eta)$ and  $H_R = H_0 \sqrt{\Omega_R} \simeq 10^{-44}\,\mathrm{GeV}$
one finds that the condition (\ref{condith}) implies that this decay channel will remain open within the window of  scale factors
\be 10^{-28}\leq a(\eta) \ll 10^{-21} \,,  \label{amin}\ee corresponding to the temperature range $10^{8}\,\mathrm{GeV} < T(t) \leq 10^{15}\mathrm{GeV}$ during the (RD) dominated era. In this temperature regime, the heavier daughter particles in this example are also  typically ultrarelativistic.

Under these circumstances the results from eqns. (\ref{URdelaw},\ref{probUR}) are valid during the time interval in which this decay channel remains open,  determined by the inequality (\ref{amin}). Eventually, however as the expansion proceeds both the local energy and expansion rate diminish and this channel closes. The detailed dynamics of this phenomenon must be studied numerically for a given range of parameters.

\begin{figure}[ht!]
\begin{center}
\includegraphics[height=4in,width=4in,keepaspectratio=true]{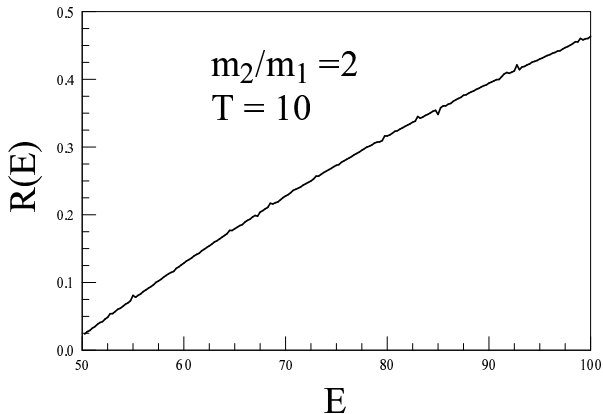}
\caption{The integral R(E) vs. E, for $m_2/m_1=2~;~\widetilde{T}=10$ in units of $m_1$.  }
\label{fig:pop}
\end{center}
\end{figure}

The integration of the convolution of the spectral density with the sine function and the further integration to obtain the decay law is extremely challenging and time consuming because of the wide separation of scales and the rapid oscillations. In a more realistic model with specific parameters such endeavor would be necessary for a detailed assessment of the contribution from the new open channels. Here we provide a ``proof of principle'' by displaying in fig. (\ref{fig:pop}) the result of the integral (see \ref{disp} and \ref{rhok0k})
\be R(E) = \int^\infty_{k^*_0}  dk_0 ~   \Bigg[  \frac{k^2_0-E^2-(4\,m^2_2-m^2_1)}{k^2_0-E^2+m^2_1} \Bigg]^{1/2} \,  \frac{\sin\Big[\big(k_0 - E  \big)  {T} \Big]}{ \big(k_0 - E  \big) } ~~;~~ k^*_0 = \sqrt{E^2+ (4m^2_2 -m^2_1)}\,  \label{RofE} \ee for $4m^2_2 > m^2_1$ so that $E$ is below threshold.

The range of $E,T$ are chosen to comply with the validity of the adiabatic condition $ET\gg 1$. This figure shows that the uncertainty ``opens'' the threshold to decaying into heavier particles, the example in  the figure corresponds to $m_2=2m_1$. We have numerically confirmed that as $T$ increases $R(E)$ diminishes as a consequence of a smaller overlap. As the scale factor increases these new decay channels close, allowing only the two body decay for $m_1 > 2m_2$ and the decay rate is given by the long time limit (\ref{gammalongtime})
\be \Gamma(\eta) = \Gamma_0 \, \frac{a(\eta)}{\gamma_k(\eta)} ~~;~~ \Gamma_0 = \frac{\lambda^2 }{8\pi\,m_1} \, \Bigg[1- \frac{4\,m^2_2}{m^2_1} \Bigg]^{1/2} \,\Theta(m^2_1-4\,m^2_2) \,, \label{gammafina}\ee where $\Gamma_0$ is the usual decay rate at rest in Minkowski space time. Following the analysis of the previous section, one now finds a decay law similar to that in eqn. (\ref{gammainter}) but with $\Gamma_0$ now given by (\ref{gammafina}).

\vspace{2mm}

\textbf{Daughters pair probability:}

\vspace{2mm}

With the solution for the amplitude of the single particle state, we can now address the amplitude for the decay products from the result (\ref{kappapop}) with $\ket{\kappa} = \ket{1^{(2)}_{\vp}, 1^{(2)}_{\vq}}$ and
$\ket{A} = \ket{1^{(1)}_{\vk}}$. The decay product is a \emph{correlated pair of daughter particles}. The corresponding matrix element is given by (\ref{mtxele}) in terms of the zeroth order adiabatic mode functions (\ref{zeromo}). Writing the solution for the decaying amplitude
\be C^{(1)}_{\vk}(\eta)= e^{-\int^\eta_{\eta_i} \mathcal{E}^{(1)}_{k}(\eta'')d\eta''}  \label{solu1}\ee where   $\mathrm{Re}\big[\mathcal{E}^{(1)}_{k}(\eta)\big] = \Gamma_k(\eta)/2$, and neglecting the contribution from the imaginary part which amounts to a renormalization of the frequencies\cite{boyww,boycosww}, we find (using \ref{kappapop})
\be C^{(2)}_{\vp,\vq}(\eta)  = -i \frac{2\lambda }{V^{1/2}}\,\int^{\eta}_{\eta_i}\,\frac{e^{i\int^{\eta'}_{\eta_i} \Big[\omega^{(2)}_{\vp}(\eta'')+ \omega^{(2)}_{\vq}(\eta'')-\omega^{(1)}_{\vk}(\eta'') \Big] d\eta'' }}{\Bigg[2\omega^{(2)}_{\vp}(\eta')\,2\omega^{(2)}_{\vq}(\eta')2\omega^{(1)}_{\vk}(\eta')\Bigg]^{1/2}}~ e^{-\int^{\eta'}_{\eta_i} \Gamma_k(\eta'')/2 ~ d\eta''} ~d\eta'~~;~~ \vq= \vk-\vp \,. \label{popuc2} \ee The time integral is extremely challenging and can only be studied numerically. We can make progress by implementing the same approximations discussed above. Since $\Gamma_k$ depends on the slowly varying frequency, it itself varies slowly, therefore we will consider an interval in $\eta$ so that the decay rate remains nearly constant,
replacing the exponentials by their lowest order expansion in $\eta'-\eta_i$. During this interval we find the following  approximate form of the daughter pair  probability,
\be  |C_{\vp,\vk}(\eta)|^2  \approx    \frac{ \lambda^2}{2\omega^{(1)}_k(\eta)\, \omega^{(2)}_p(\eta)\,\omega^{(2)}_q(\eta) \, V }  ~~  \frac{ \Bigg| 1-e^{-\Gamma_k(\eta) \eta/2}\,e^{-i \big(\omega^{(1)}_k(\eta)-\omega^{(2)}_p(\eta)-\omega^{(2)}_q(\eta) \big)\eta }\Bigg|^2}{ \big(\omega^{(1)}_k(\eta)-\omega^{(2)}_p(\eta)-\omega^{(2)}_q(\eta) \big)^2 +\frac{\Gamma^2_k(\eta)}{4}  }~~;~~ \vq = \vk-\vp \,, \label{popuc2ap}\ee where we set $\eta_i=0$. This expression is only valid in restricted time interval, its main merit is that it agrees with the result in Minkowski space time (see appendix \ref{app:Mink}) and describes the early build up of the daughters population from the decay of the parent particle. The occupation number of daughter particles   is obtained by calculating the expectation value of the number operators $a^\dagger_{\vq} a_{\vq}~;~a^\dagger_{\vp} a_{\vp}$ in the time evolved state, it is straightforward to find
\be \langle a^\dagger_{\vq} a_{\vq} \rangle = \langle a^\dagger_{\vp} a_{\vp} \rangle = |C_{\vp,\vk}(\eta)|^2 \,,\label{occunum}\ee the fact that these occupation numbers are the same is a consequence of the pair correlation.

A more detailed assessment of the population build up and asymptotic behavior requires a full numerical study for a range of parameters.

\section{Discussion}

There are several aspects and results of this study that merit further discussion.

\vspace{1mm}

\textbf{Spontaneous vs. stimulated decay:} We have focused on the dynamics of decay from an initial state assuming that there is no established population of daughter particles in the plasma that describes an (RD) cosmology. If there is such population there is a contribution from \emph{stimulated} decay in the form of extra factors $1+n$ for each bosonic final state where $n$ is the occupation of the particular state. These extra factors enhance the decay. On the other hand, if the particles in the final state are fermions (a case not considered in this study), the final state factors are $1-n$ for each fermionic daughter species and the decay rate would decrease as a consequence of Pauli blocking. The effect of an established population of daughter particles on the decay rate   clearly merits further study.

 \vspace{1mm}

\textbf{Medium corrections:} In this study we focus on the corrections to the decay law arising solely from the cosmological expansion as a prelude to a more complete treatment of kinetic processes in the early Universe. In this preliminary study we have not included the effect of medium corrections  to the interaction vertices or masses. Finite temperature effects, and in particular in the early radiation dominated stage, modify the effective couplings and masses, for example a Yukawa coupling to fermions or a bosonic quartic self interaction would yield finite temperature corrections to the masses $\propto T^2$. These modifications may yield important corrections to the spectral densities  and may also modify threshold kinematics. However, the \emph{dynamical} effects such as threshold relaxation, consequences of uncertainty and delayed decay (relaxation) as a consequence of cosmological redshift of time dilation are robust phenomena that do not depend on these aspects. Our formulation applies to the time evolution of (pure) states. In order to study the time evolution of distribution functions it must be extrapolated to the time evolution of a density matrix, from which one can extract the quantum kinetic equations  including the effects of cosmological expansion described here. This program merits a deeper study beyond the scope of this article. We are currently pursuing several of these aspects.

\vspace{1mm}

\textbf{Cosmological particle production:}  Our study has focused on the zeroth adiabatic order as a prelude to a more comprehensive program. We have argued that at the level of the Hamiltonian, the creation and annihilation operators introduced in the quantization procedure create and destroy particles as identified at leading adiabatic order and diagonalize the Hamiltonian at leading (zero) order. Beyond the leading order, there emerge contributions that describe the creation (and annihilation) of \emph{pairs} via the cosmological expansion. We have argued that these processes are of higher order in the adiabatic expansion, therefore can be consistently neglected to leading order. For weak coupling, including these higher order processes of cosmological particle production (and annihilation) in the calculation of the\emph{ decay rate} (and decay law) will result in higher order corrections to the rate of the form $\lambda^2 \times \,(\mathrm{higher}~\mathrm{order}~\mathrm{adiabatic})$. However, once these processes are included \emph{at tree level}, namely at the level of \emph{free field} particle production, they may actually \emph{compete} with the decay process. It is   possible that for weak coupling, cosmological particle production (and annihilation) competes on similar time scales with decay, thereby perhaps ``replenishing'' the population of the decaying particle. The study of these competing effects requires the equivalent of a quantum kinetic description including the gain from particle production and the loss from decay (and absorption of particles into the vacuum). Such study will be the focus of a future report.

\vspace{1mm}

\textbf{Validity of the adiabatic approximation:} The adiabatic approximation relies on the ratio $H(t)/E_k(t) \ll 1$ (\ref{adpara}). In a radiation dominated cosmology the Hubble radius ($H^{-1}(t)$) grows as $a^2(t)$ and during matter domination it grows as $a^{3/2}(t)$ whereas physical wavelengths grow as $a(t)$, with $a(t)$ the scale factor. During these cosmological eras,   physical wavelengths become deeper inside the Hubble radius and the ratio $H(t)/E_k(t)$ diminishes fast. Therefore if the condition $H(t)/E_k(t)\ll 1$ is satisfied at the very early stages during radiation domination,  its validity \emph{improves} as the cosmological expansion proceeds.

\vspace{1mm}

\textbf{Modifications to BBN?} The results obtained in the previous sections show potentially important modifications to the decay law during the (RD) cosmological era. An important question is whether these corrections affect standard BBN. To answer this question we focus on neutron decay, which is an important ingredient in the primordial abundance of Helium and heavier elements. The neutron is ``born'' after the QCD confining phase transition at $T_{QCD} \simeq 150 \,\mathrm{MeV}$ at a time $t_{QCD} \simeq 10^{-5}\,s$ hence neutrons are ``born'' non-relativistically. With a mass $M_N \simeq 1\,\mathrm{GeV}$ and a typical physical energy $\simeq T_{QCD}$ the transition time $t_{nr} \simeq 10^{-6}\,s \simeq t_{QCD}$. The neutron's lifetime $\simeq 900 \,s$ implies that $\Gamma_0 \,t_{nr}/2 \simeq 10^{-9}$ and the modifications from the decay law determined by the extra factor in (\ref{powlaw}) are clearly irrelevant. Therefore it is not expected that the modifications of the decay law found in the previous sections would affect the dynamics of BBN and the primordial abundance of light elements. There is, however, the possibility that \emph{other}   degrees of freedom, such as, sterile neutrinos for example,  whose decay may inject energy into the plasma with potential implications for  BBN. Such a possibility has been raised in refs.\cite{bbnconstdecay}-\cite{salvatibbn} with regard to the abundance of $^{7}Li$. The decay law of these other species of particles (such as sterile neutrinos beyond the standard model) \emph{could} be modified and their efficiency for energy injection and potential impact on BBN may be affected by these modifications. Such possibility remains to be studied.

\vspace{1mm}

\textbf{Wave packets:} We have studied the decay dynamics from an initial state corresponding to a single particle state with a given comoving wavector. However, it is possible that the decaying parent particle is not created (``born'') as a single particle eigenstate of momentum, but in a wave packet superposition. Taking into account this possibility is straightforward within the Wigner-Weisskopf method, and it has been considered in Minkowski space time in ref.\cite{boyww}. Consider an initial wave packet as a linear superposition of single particle states of the parent field, namely $\ket{\mathbf{1}^{(1)}}= \sum_{\vk} C^{(1)}_{\vk}(\eta_i)\ket{1^{(1)}_{\vk}}$, where $C^{(1)}_{\vk}(\eta_i)$ are the Fourier coefficients of a wavepacket localized in space (for example a Gaussian wave-packet). Implementing the Wigner-Weisskopf method, the time evolution of this state leads to  the solution (\ref{CAfina}) for the coefficients  with $C_A(\eta_i)  = C^{(1)}_{\vk}(\eta_i)$, and by Fourier transform one obtaines the full space-time evolution of the wavepacket\cite{boyww}.  Such an extension presents no conceptual difficulty, however, the major technical complication would be to extract the decay law: as pointed out in the previous section, the main difference with the result in Minkowski space time is that the time dilation factors depend explicitly on time through the cosmological redshift. In a wave packet description, each different wavector component features a different time dilation factor with a differential red-shift between the various components. This will modify the evolution dynamics in several important ways: there is spreading associated with dispersion, the different time dilation factors for each wavevector imply a superposition of different decay time scales, and finally, each different time dilation factor features a different time dependence through the cosmological redshift. All these aspects amount to important technical complexities that merit further study.

\vspace{1mm}

\textbf{Caveats:} The main approximation invoked in this study, the adiabatic approximation, relies on the physical wavelength of the particle to be deep inside the physical particle horizon at any given time, namely, much smaller than the Hubble radius. If the decaying parent particle is produced (``born'') satisfying this condition, this approximation becomes more reliable with cosmological expansion as  the Hubble radius grows faster than a physical wavelength during an (RD) or (MD) cosmology. However, it is possible that such particle has been produced during the inflationary, near de Sitter  stage, in which case the Hubble radius remains nearly constant and the physical wavelength is stretched beyond it. In this situation, the adiabatic approximation as implemented in this study breaks down. While the physical wavelength remains outside the particle horizon, the evolution must be obtained by solving the equations of motion for the mode function.  During the post inflationary evolution well after the physical wavelength of the parent particle  re-enters the Hubble radius  the adiabatic approximation becomes reliable. However, it is possible that while the physical wavelength is \emph{outside} the particle horizon during (RD) (or (MD)) the parent particle has decayed substantially  with the ensuing growth of the daughter population. The framework developed in this study would need to be modified to include this possibility, again a task beyond the scope and goals of this article.

\section{Conclusions and further questions}

Motivated by the phenomenological importance of particle decay in cosmology for physics within and beyond the standard model, in this article we initiate a program to provide a systematic framework to obtain the \emph{decay law} in the standard post inflationary cosmology. Most of the treatments of phenomenological consequences of particle decay in cosmology describe these processes in terms of a decay rate obtained via usual S-matrix theory in Minkowski space time. Instead, recognizing that rapid cosmological expansion may modify this approach with potentially important phenomenological consequences, we study particle decay by combining a physically motivated adiabatic expansion and a \emph{non-perturbative} quantum field theory method which is an extension of the ubiquitous Wigner-Weisskopf theory of atomic line widths in quantum optics\cite{book1}. The adiabatic expansion relies on a wide separation of scales: the typical wavelength of a particle is much smaller than the particle horizon (proportional to the Hubble radius) at any given time. Hence we introduce the \emph{adiabatic} ratio $H(t)/E_k(t)$ where $H(t)$ is the Hubble rate and $E_k(t)$ the (local) energy measured by a comoving observer. The validity of the adiabatic approximation relies on $H(t)/E_k(t) \ll 1$ and is fulfilled under   \emph{most} general circumstances of particle physics processes in cosmology.

The Wigner-Weisskopf framework allows to obtain the survival probability and  \emph{decay law} of a parent particle along with the   probability of population build-up for the daughter particles (decay products).  We implement this framework within a model quantum field theory to study the generic aspects of particle decay in an expanding cosmology, and compare the results of the cosmological setting with that of Minkowski space time.

One of our main results is a \emph{cosmological Fermi's Golden Rule} which  features an energy uncertainty determined by the particle horizon ($\propto 1/H(t)$) and yields the\emph{ time dependent decay rate}. In this study we obtain two main results: i) During the (RD) stage, the survival probability of the decaying (single particle) state  may be written in terms of an \emph{effective time dependent rate} $\widetilde{\Gamma}_k(t)$   as $P(t) = e^{-\widetilde{\Gamma}_k(t)\,t}$. The effective rate is characterized by a time scale $t_{nr}$ (\ref{gammainter}) at which the particle transitions from the relativistic regime ($t \ll t_{nr}$) when $P(t) = e^{-(t/t^*)^{3/2}}$  to the non-relativistic regime ($t\gg t_{nr}$) when $P(t)= e^{-\Gamma_0\,t}\, \Big(\frac{t}{t_{nr}}\Big)^{\Gamma_0 t_{nr}/2}$ where $\Gamma_0$ is the Minkowski space-time decay width at rest. Generically the decay is \emph{slower} in an expanding cosmology than in Minkowski space time. Only for a particle that has been produced (``born'') at rest in the comoving frame is the decay law  asymptotically the same as in Minkowski space-time. Physically the reason for the delayed decay is that for non-vanishing momentum   the decay rate features the (local) time dilation factor, and in an expanding cosmology the (local)  Lorentz factor depends on time through the  cosmological redshift. Therefore lighter particles that are produced with a large Lorentz factor decay with  an effective longer lifetime. ii) The second, unexpected result of our study is a \emph{relaxation of thresholds} as a consequence of the energy uncertainty determined by the particle horizon. A distinct consequence of this uncertainty  is the opening of new decay channels to decay products that are \emph{heavier} than the parent particle. Under the validity of the adiabatic approximation, this possibility is available when $2\pi E_k(t) H(t) \gg 4m^2_2-m^2_1$ where $m_1,m_2$ are the masses of the parent, daughter particles respectively. As the expansion proceeds this channel closes and the usual kinematic threshold constrains the phase space available for decay. Both these results \emph{may} have important phenomenological consequences in baryogenesis, leptogenesis, and dark matter abundance and constraints which remain to be studied further.

\vspace{1mm}

\textbf{Further questions:}

We have focused our study on a simple quantum field theory model that is not directly related to the standard model of particle physics or beyond. Yet, the results have a compelling and simple physical interpretation that is likely to transcend the particular model. However, the analysis of this study must be applied to other fields in particular fermionic degrees of freedom and vector bosons. Both present new and different technical challenges primarily from their couplings to gravity which will determine not only the scale factor dependence of vertices but also the nature of the mode functions (spinors in particular). As mentioned above, cosmological particle production is not included to leading order in the  adiabatic approximation  but must be   consistently included beyond leading order. The results of this study point to  interesting avenues to pursue further: in particular the relaxation of kinematic thresholds from the cosmological uncertainty opens the possibility for unexpected phenomena and possible modifications to processes, such as inverse decays, the dynamics of thermalization and detailed balance. These are all issues that merit a deeper study, and we expect to report on some of them currently in progress.

\acknowledgments DB gratefully acknowledges support from the US National Science
Foundation through grant PHY-1506912. The work of ARZ is supported in part by the
US National Science Foundation through grants AST-1516266 and AST-1517563 and
by the US Department of Energy through grant DE-SC0007914.

\appendix

\section{Particle Decay in Minkowski Spacetime}\label{app:Mink}

In order to understand more clearly the decay law in cosmology, it proves convenient to study the decay of a massive particle into two   particles in Minkowski space time implementing the Wigner-Weisskopf method.

\vspace{1mm}

\textbf{Integrating in momentum first: massless daughters}

This is achieved from the expression (\ref{sigcos}) by simply taking
\be \eta \rightarrow t~~;~~ a(\eta) \rightarrow  1~~,~~ g^{(1)}_k(\eta) \rightarrow \frac{e^{-iE_k\,t}}{\sqrt{2E_k}} ~~;~~ g^{(2)}_k(\eta) \rightarrow \frac{e^{-ik\,t}}{\sqrt{2k}} \,, \label{mink}\ee with $E_k = \sqrt{k^2+m^2}$, leading to
\be \Sigma_k(t-t')= \frac{\lambda^2}{E_k}\, \int \frac{d^3p}{(2\pi)^3} ~~\frac{e^{i(E_k-p-q)(t-t')}}{2p\,2 q} ~~;~~ q = |\vk-\vp|\,. \label{sigmink}\ee The integral over $p$ can be done by writing $d^3 p = p^2 dp \, d(\cos(\theta))$ and changing variables from $\cos(\theta)$ to $q = \sqrt{k^2+p^2-2kp\cos(\theta)}$ with $d(\cos(\theta))/q  = -dq/k\,p$, and introducing a convergence factor $t-t' \rightarrow (t-t'-i\epsilon)$ with $\epsilon \rightarrow 0^+$. We find
\be \Sigma_k(t-t')=  \frac{-i\,\lambda^2}{16\pi^2\,E_k}~\frac{e^{i(E_k-k)(t-t')}}{(t-t'-i\epsilon)} = \frac{ \lambda^2}{16\pi^2\,E_k}~ e^{i(E_k-k)(t-t')} \,\Bigg[ -i\,\mathcal{P}\Bigg(\frac{1}{t-t'} \Bigg) + \pi\,\delta(t-t')\Bigg] \,, \label{sigminkfin}\ee  and
\be \mathrm{Re}\Sigma_k(t-t') =  \frac{ \lambda^2}{16\pi^2\,E_k}~\Bigg\{ \pi\,\delta(t-t') + \frac{\sin\big[(E_k-k)(t-t') \big]}{(t-t')}\Bigg\}\,.\label{Resig}\ee
This expression yields a \emph{time dependent} decay rate $\Gamma(t)$ given by
\be \Gamma(t) = 2 \int^t_0 \mathrm{Re}\Sigma_k(t-t') \,dt' = \frac{\lambda^2}{8\pi\,E_k}\,\frac{1}{2}\Big[1+\frac{2}{\pi}\,Si[(E_k-k)t] \Big] \,,\label{gammaoft}\ee where $Si[x]$ is the sine-integral function with asymptotic limit  $Si[x] \rightarrow \pi/2$ for $x\rightarrow \infty$. The time scale to reach the asymptotic behavior
\be t_{asy} \propto \frac{1}{E_k-k}\,, \label{tasy}\ee therefore the approach to asymptotia and to the full width takes a much longer time for an ultrarelativistic particle with $t_{asy} \propto 2k/m^2$, whereas it is much shorter in the non-relativistic case $t_{asy} \propto 1/m$. In S-matrix theory in Minkowski space time one takes $t\rightarrow \infty$, and obviously in this limit the $Si-$ function reaches its asymptotic value, therefore  the time dependence of the rate cannot be gleaned.

\vspace{1mm}

\textbf{Integrating in time first: massive particles and Fermi's Golden rule.}

Let us consider now the full dispersion relations for the daughter particles, calling $E_k$ that of the parent decaying particle and $\omega_p = \sqrt{p^2+m^2_2}$ that of the daughter.
From (\ref{Evar}) and (\ref{probaA}), we need
\be \mathcal{E}_k[t;t] = \int^t_0 \Sigma_k(t-t')\,dt' ~~;~~ \Gamma_k(t) = 2\,\mathrm{Re}\mathcal{E}_k[t,t]\,.  \label{Wmin}\ee
We find
\be \Gamma_k(t) = \frac{2\,\lambda^2}{E_k}\, \int \frac{d^3p}{(2\pi)^3} ~~\frac{\sin\big[(E_k -\omega_p-\omega_q)\,t\big] }{2\omega_p\,2 \omega_q\,\big[(E_k -\omega_p-\omega_q)\big]} ~~;~~ q = |\vk-\vp|\,, \label{gamink}\ee the asymptotic long time limit
\be \frac{\sin\big[(E_k -\omega_p-\omega_q)\,t\big] }{ \big[(E_k -\omega_p-\omega_q)\big]}   ~~ ~~ \overrightarrow{t \rightarrow \infty} ~~~~ \pi\,\delta \big(E_k -\omega_p-\omega_q \big) \,,\label{FGR}\ee yields

\be \Gamma_k(t) ~~{}_{\overrightarrow{t \rightarrow \infty}}  ~~~\frac{ \lambda^2}{ E_k}\, \int \frac{d^3p}{(2\pi)^3\,2\omega_p\,2 \omega_q} ~~ (2\pi)\, \delta \big(E_k -\omega_p-\omega_q \big) \,,\label{gamafgr} \ee this is simply Fermi's Golden rule which yields the standard result for the decay rate
\be \Gamma_k = \frac{\lambda^2}{8\pi\,E_k} \, \Bigg[1- \frac{4\,m^2_2}{E^2_k-k^2} \Bigg]^{1/2} \,\Theta(E^2_k - k^2-4\,m^2_2)\,. \label{Gamaminko}\ee Although $E^2_k-k^2 = m^2_1$ we have left the result in the form shown to make use of it in the cosmological case and to highlight the threshold.

Before taking the limit $t\rightarrow \infty$ the real time rate (\ref{gamink}) can be conveniently written in a dispersive form, namely

\be \Gamma_k(t) = \int^{\infty}_{-\infty} \rho(k_0,k)\, \frac{\sin\big[(k_0-E_k)\,t\big] }{ \big[\pi\,(k_0-E_k )\big]}  \, dk_0 \label{dispermink}\ee with the spectral density
\be \rho(k_0,k) = \frac{ \lambda^2}{E_k}\, \int \frac{d^3p}{(2\pi)^3} ~~\frac{(2\pi)\,\delta(k_0 -\omega_p-\omega_q) }{2\omega_p\,2 \omega_q} ~~;~~ q = |\vk-\vp|\,, \label{rho}\ee which, following the steps leading up to (\ref{Gamaminko}) is given by
\be \rho(k_0,k) =\frac{\lambda^2}{8\pi\,E_k} \, \Bigg[1- \frac{4\,m^2_2}{k^2_0-k^2} \Bigg]^{1/2} \,\Theta(k^2_0 - k^2-4\,m^2_2)\,\Theta(k_0)  \,. \label{rhofin}\ee

The case of massless daughter's particles $m_2=0$ is particularly simple, yielding
\be \Gamma_k(t) = \frac{\lambda^2}{8\pi^2\,E_k} \int^{\infty}_{-(E_k-k)t}\,\frac{\sin(x)}{x} ~ dx = \frac{\lambda^2}{8\pi\,E_k}\,\frac{1}{2}\Big[ 1+ \frac{2}{\pi} Si[(E_k-k)t] \Big]\,. \label{promink}\ee This expression of course agrees with eqn. (\ref{gammaoft}) and clarifies  the emergence of a prompt term given by $\delta(t-t')$ in (\ref{sigminkfin}) and the ``rising'' term, namely the $Si$ function that reaches its asymptotic value $\pi/2$ over a time scale $\approx 1/(E_k-k)$, by   integrating in time first.

Using the result (\ref{kappapop}) adapted to Minkowski space time, with the state $\ket{\kappa} = \ket{1^{(2)}_{\vp},1^{(2)}_{\vq}}$ the amplitude for daughter particles becomes
\be C_{\vp,\vk}(t)  =  -i\bra{1^{(2)}_{\vp}1^{(2)}_{\vq}}H_I \ket{1^{(2)}_{\vk}} \int^{t}_0 e^{-i \big(E_k-\omega_p-\omega_q \big)t' }\, e^{-\Gamma_k t'/2}\,dt' \label{ampdaumink}\ee with the probability given by
\be  |C_{\vp,\vk}(t)|^2  =   \frac{\lambda^2}{2\,\omega^{(1)}_k\, \omega^{(2)}_p\,\omega^{(2)}_q \, V }  \,  \frac{\Big| 1-e^{-\Gamma_k t/2}\,e^{-i \big(E_k-\omega_p-\omega_q \big)t }\Big|^2}{\Big[ (E_k-\omega_p-\omega_q )^2+\frac{\Gamma^2_k}{4}  \Big]}~~;~~ \vq = \vk-\vp \,. \label{popdaumink}\ee

\vspace{2mm}

\section{First order adiabatic correction for massive daughters.}\label{app:validity}

There are two contributions of first adiabatic order in the time  integrals up to $\eta$ of  equation (\ref{sigadzero}): 1) keeping the quadratic term $(\eta-\eta')^2$ multiplied by derivatives of the frequencies in the exponential (see eqn. (\ref{adex})). With the substitution $\tau = \omega^{(1)}_k(\eta)\,(\eta-\eta')$ this term is proportional to $\tau^2$,  and 2) in the first order expansion of the scale factor and the frequencies obtained from the expression (\ref{poftau}), this term is proportional to $\tau$. Both terms are of first adiabatic order, hence are multiplied by $H(t)/E_k(t)  \equiv 1/z$ where we have taken the frequency of the parent particle as reference frequency. The contributions to the integral (here we set $\eta_i=0$)
$$  \int_{0}^{\eta} \Sigma_k(\eta,\eta')\,d\eta'$$ are of the form
$$ \frac{1}{z}   \int^{z}_0 (a\,\tau + i\,b \,\tau^2) \,e^{i\Big[1-\frac{\omega^{(2)}_p(\eta)}{\omega^{(1)}_k(\eta)}-
\frac{\omega^{(2)}_q(\eta)}{\omega^{(1)}_k(\eta)}\Big]\tau} \, d\tau $$ where $a,b$ are z-independent  coefficients but depend on the momenta. Introducing the dispersive form of the momentum integrals as in equation (\ref{disp}) and introducing
\be \epsilon = \frac{k_0-E^{(1)}_k}{E^{(1)}_k}\,,
 \label{epsi}\ee
 we find the following contributions to the corrections to $\mathrm{Re}\Sigma_k$:
 \be \mathrm{Re} \int^{z}_0 \tau e^{i\epsilon \tau} d\tau = f_1(\epsilon,z)= \frac{d}{d\epsilon} \Big[\frac{(1-\cos(\epsilon\,z)}{\epsilon} \Big]\label{inte1}\ee

\be \mathrm{Re} \int^{z}_0 i\, \tau^2 e^{i\epsilon \tau} d\tau = f_2(\epsilon,z) = \frac{d^2}{d\epsilon^2} \Big[\frac{(1-\cos(\epsilon\,z)}{\epsilon} \Big]\,.\label{inte2}\ee Changing integration variables from $k_0$ to $\epsilon$ in the dispersive form and writing the spectral density $\rho(k_0,k) \equiv \rho(\epsilon) $ to simplify notation the corrections to the rate $\Gamma_k(\eta)$ to first adiabatic order are determined by the following integrals
\be I_{1,2}(z)= \frac{1}{z} \int^{\infty}_{-\infty} \rho(\epsilon) f_{1,2}(\epsilon,z)\,d\epsilon\, \,,\label{I12ofz} \ee for comparison, in terms of the same variables,  the zeroth order adiabatic term is given by
\be I_0(z) = \int^{\infty}_{-\infty} \rho(\epsilon) \frac{\sin(\epsilon z)}{\epsilon}\,d\epsilon\,.  \label{I0ofz}\ee

\begin{figure}[ht!]
\begin{center}
\includegraphics[height=4in,width=4in,keepaspectratio=true]{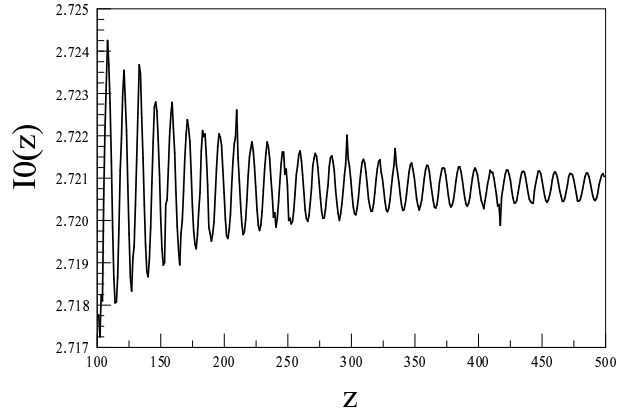}
\caption{The integral $I_0(z)$ vs. z, for $m_2/m_1=0.25~,~k=0$ .  }
\label{fig:izero}
\end{center}
\end{figure}

\begin{figure}[ht!]
\begin{center}
\includegraphics[height=3.5in,width=3.2in,keepaspectratio=true]{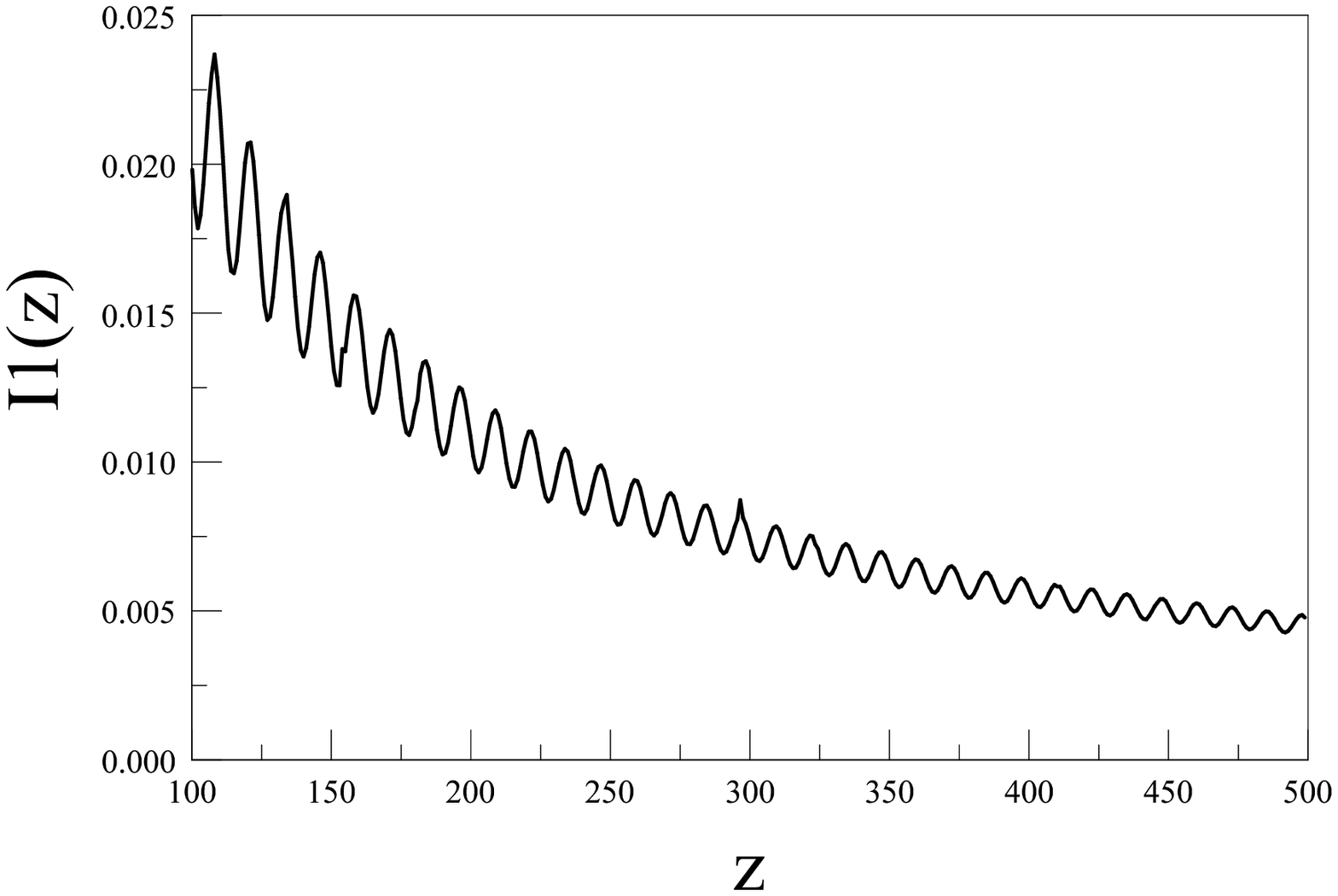}
\includegraphics[height=3.5in,width=3.2in,keepaspectratio=true]{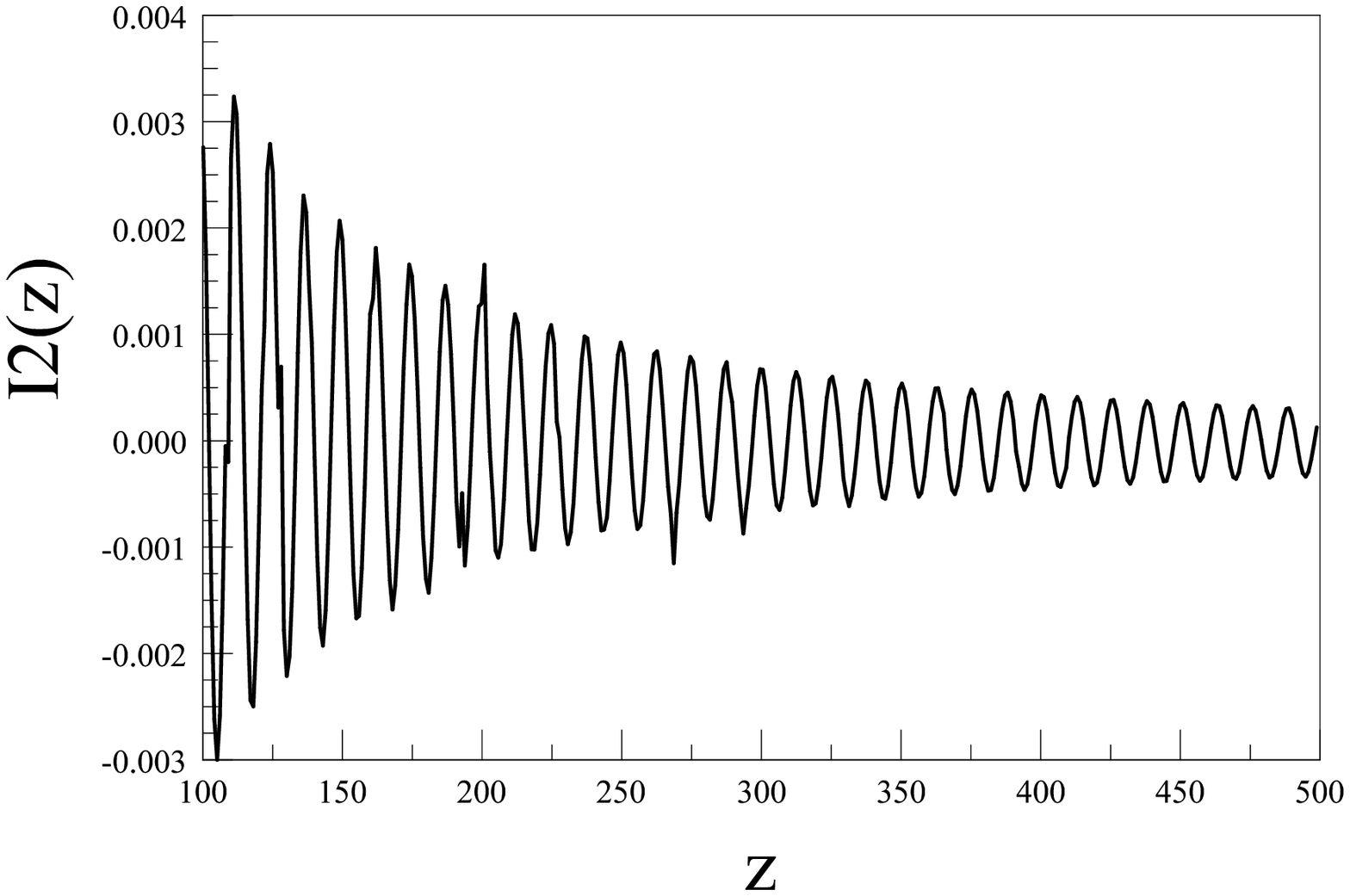}
\caption{The integrals $I_1(z),I_2(z)$ vs. z, for $m_2/m_1=0.25~,~ k=0$.  }
\label{fig:iunodos}
\end{center}
\end{figure}

The function $f_0(\epsilon,z)=\sin(\epsilon z)/\epsilon$ is the usual function of Fermi's Golden Rule: for large $z$  it is sharply localized near $\epsilon \simeq 0$ with total area $= \pi$, it becomes a delta function in the large $z$ limit, probing the region $\epsilon \simeq 0$ of the spectral density. The function
$f_1(\epsilon,z)$ is even in $\epsilon$ and for large $z$ is also  localized near $\epsilon \simeq 0$ but in this limit it becomes the \emph{difference} of delta functions multiplied by $z$ plus subdominant terms. Because this function is a total derivative  the total integral area is independent of $z$ and vanishes in  the integration  domain $-\infty < \epsilon < \infty$. If $m_1$ is above the threshold the total integral does not vanish but becomes independent of $z$ and small as $z\rightarrow \infty$, thus we expect $I_1(z)$ to fall off rapidly with $z$. Finally, the function $f_2(\epsilon,z)$ is odd in $\epsilon$ and for large $z$ is also localized near $\epsilon \simeq 0$ but vanishing at $\epsilon =0$ and rapidly varying in this region, averaging out the integral over the spectral density. Thus we also expect that $I_2(z)$ falls off with $z$ with nearly zero average because of being odd in $\epsilon$. Figures (\ref{fig:izero}, \ref{fig:iunodos}) display $I_0,I_1,I_2$ for a representative set of parameters. The main features are confirmed by a comprehensive numerical study for a wide range of parameters for $m_1 > 2m_2$ (above threshold). If $m_1$ is below the two particle threshold, the spectral density vanishes in the region of support of the functions $f_1,f_2$ thereby yielding   rapidly vanishing integrals for large $z$. We have confirmed numerically that both $I_1,I_2$ vanish very rapidly as a function of $z$ in this case, remaining perturbatively small when compared to $I_0$. Therefore this   study confirms that the first order adiabatic corrections are indeed subleading as compared to the leading (zeroth) order contribution for large $z= E_k(t)/H(t)$.

\end{document}